\documentclass[twocolumn]{aastex701}
\usepackage{glossaries}

\newcommand{\SNIa}{SN\,Ia }
\newcommand{\SNeIa}{SNe\,Ia }
\newcommand{\SNIanospace}{SN\,Ia}
\newcommand{\SNeIanospace}{SNe\,Ia}
\usepackage{float}
\usepackage{gensymb}
\usepackage{comment}

\newcommand{\SSci}{Astrophysics \& Space Institute, Schmidt Sciences, New York, NY 10011, USA}
\newcommand{\PPLLC}{Project Pearl, Schmidt Sciences, New York, NY 10011, USA}
\newcommand{\UAA}{Department of Astronomy and Steward Observatory, University of Arizona, Tucson, AZ 85721, USA}
\newcommand{\UCB}{University of California, Berkeley, CA 94720, USA}
\newcommand{\LBL}{Lawrence Berkeley National Laboratory, Berkeley, CA 94720, USA}
\newcommand{\SSL}{Space Sciences Laboratory, University of California, Berkeley, Berkeley, CA 94720, USA}
\newcommand{\KU}{Department of Physics and Astronomy, University of Kansas, Lawrence, KS 66045, USA}
\newcommand{\UMD}{Department of Astronomy, University of Maryland, College Park, MD 20742, USA}
\newcommand{\TTU}{Department of Physics \& Astronomy, Texas Tech University, Box 41051, Lubbock, TX 79409-1051, USA}
\newcommand{\STScI}{Space Telescope Science Institute, 3700 San Martin Dr, Baltimore, MD 21218, USA}
\newcommand{\Lyon}{Universite Claude Bernard Lyon 1, CNRS, IP2I Lyon / IN2P3, IMR 5822, F-69622 Villeurbanne, France}

\usepackage{color}
\begin{document}
\title{The Lazuli Space Observatory: Architecture \& Capabilities}

%==========================

\author[0000-0001-8127-5775]{Arpita Roy}
\affiliation{\SSci}
\email{}

\author[0000-0003-0963-3640]{Stuart Feldman}
\affiliation{\SSci}
\email{}

\author[]{Peter Klupar}
\affiliation{\PPLLC}
\email{}

\author[]{John DiPalma}
\affiliation{\PPLLC}
\email{}

\author[0000-0002-4436-4661]{Saul Perlmutter}
\affiliation{\UCB}
\affiliation{\LBL}
\email{}

\author[0000-0002-0813-4308]{Ewan S. Douglas}
\affiliation{\UAA}
\email{}

\author[]{Greg Aldering}
\affiliation{\LBL}
\email{}

\author[0000-0001-8467-9767]{Gabor Furesz}
\affiliation{\SSci}
\email{}

\author[0000-0003-3715-8138]{Patrick Ingraham}
\affiliation{\UAA}
\email{}

\author[0000-0001-7409-5688]{Gudmundur Stefansson}
\affiliation{\SSci}
\affiliation{Anton Pannekoek Institute for Astronomy, University of Amsterdam, Science Park 904, 1098 XH Amsterdam, The Netherlands}
\email{}

\author[]{Douglas Kelly}
\affiliation{\UAA}
\email{}

\author[]{Fan Yang Yang}
\affiliation{\PPLLC}
\email{}

\author[0000-0002-4043-9400]{Thomas Wevers}
\affiliation{\SSci}
\email{}

\author[0000-0003-2631-5265]{Nicole Arulanantham}
\affiliation{\SSci}
\email{}

\author[0000-0003-2999-4873]{James Lasker}
\affiliation{\SSci}
\email{}

\author[0000-0002-8121-2560]{Mickael Rigault}
\affiliation{\Lyon}
\email{}

\author[0000-0001-8291-6490]{Everett Schlawin}
\affiliation{\SSci}
\email{}

\author[0000-0002-3548-4696]{Sander R. Zandbergen}
\affiliation{\PPLLC}
\email{}

\author[0000-0001-8834-9108]{S. Pete Worden}
\affiliation{\PPLLC}
\email{}

% --------------------
% Tier 2 + Contributory (alphabetical by last name)
% --------------------

\author[0000-0002-4989-6253]{Ramya Anche}
\affiliation{\UAA}
\email{}

\author[]{Heejoo Choi}
\affiliation{\UAA}
\email{}

\author[]{Ian J.\ M.\ Crossfield}
\affiliation{\KU}
\email{}

\author[]{Kevin Derby}
\affiliation{\UAA}
\email{}

\author[0009-0002-2419-8819]{Jerry Edelstein}
\affiliation{\SSL}
\email{}

\author[]{Mike Eiklenborg}
\affiliation{\UAA}
\email{}

\author[0000-0003-3703-5154]{Suvi Gezari}
\affiliation{\UMD}
\email{}

\author[]{Paul Giuliano}
\affiliation{\PPLLC}
\email{}

\author[0000-0001-9994-2142]{Justin Hom}
\affiliation{\UAA}
\email{}

\author[0000-0001-9664-0560]{Taylor J. Hoyt} \affiliation{\UCB}
\email{}

\author[0000-0002-2224-1353]{Hyukmo Kang}
\affiliation{\UAA}
\email{}

\author[0000-0002-1122-8727]{Daewook Kim}
\affiliation{\UAA}
\email{}

\author[0009-0000-4830-1484]{Keerthi Kunnumkai}
\affiliation{Department of Physics, Carnegie Mellon University, Pittsburgh, PA 15213, USA}
\email{}

\author[0000-0003-0629-5746]{Leander Lacroix}
\affiliation{\Lyon}
\email{}

\author[0000-0002-2346-3441]{Jared R. Males}
\affiliation{\UAA}
\email{}

\author[]{Thomas J. Maccarone}
\affiliation{\TTU}
\email{}

\author[0000-0001-9204-2299]{Kian Milani}
\affiliation{\UAA}
\email{}

\author[0009-0005-2445-7544]{Timothy N Miller}
\affiliation{\SSL}
\email{}

\author[]{Kelsey Lynn Miller}
\affiliation{\UAA}
\email{}

\author[]{Pierre Nicolas}
\affiliation{\UAA}
\email{}

\author[0000-0002-6011-0530]{Antonella Palmese}
\affiliation{Department of Physics, Carnegie Mellon University, Pittsburgh, PA 15213, USA}
\email{}

\author[0009-0000-2075-1109]{Jason Pero}
\affiliation{\KU}
\email{}

\author[0000-0003-3818-408X]{Laurent Pueyo}
\affiliation{\STScI}
\email{}

\author[]{Stephanie Rinaldi}
\affiliation{\UAA}
\email{}

\author[0000-0003-4102-380X]{David J. Sand}
\affiliation{\UAA}
\email{}

\author[0000-0002-5094-2245]{Christian Schneider}
\affiliation{ITAP, Kiel University, Leibnizstrasse 15, 24118 Kiel, Germany}
\email{}

\author[0000-0002-8780-8226]{Sanchit Sabhlok}
\affiliation{\UAA}
\email{}

\author[0000-0002-3957-2474]{Arfon Smith}
\affiliation{\SSci}
\email{}

\author[]{Irina I. Stefan}
\affiliation{\UAA}
\email{}

\author[]{Saraswathi Kalyani Subramanian}
\affiliation{\UAA}
\email{}

\author[]{Kyle Van Gorkom}
\affiliation{\UAA}
\email{}

\author[0009-0008-2893-774X]{Andre F. Wong}
\affiliation{\UAA}
\email{}

\author[0009-0002-0291-4933]{Jaegun Yoo}
\affiliation{\UAA}
\email{}

\author[0000-0002-3696-2127]{Md Abdullah Al Zaman}
\affiliation{\KU}
\email{}

% \author[*]{The Lazuli science team}
% \affil{}
% \email{}

% \collaboration{all}{Lazuli Science Team}

\begin{abstract}
The Lazuli Space Observatory is a 3-meter aperture astronomical facility designed for rapid-response observations and precision astrophysics across visible to near-infrared wavelengths (400--1700~nm bandpass). An off-axis, freeform telescope delivers diffraction-limited image quality (Strehl $>$0.8 at 633~nm) to three instruments across a wide, flat focal plane. The three instruments provide complementary capabilities: a Wide-field Context Camera (WCC) delivers multi-band imaging over a $35'\times12'$ footprint with high-cadence photometry; an Integral Field Spectrograph (IFS) provides continuous 400--1700~nm spectroscopy at $R\sim100$--500 for stable spectrophotometry; and an ExtraSolar Coronagraph (ESC) enables high-contrast imaging expected to reach raw contrasts of $10^{-8}$ and post-processed contrasts approaching $10^{-9}$. Operating from a 3:1 lunar-resonant orbit, Lazuli will respond to targets of opportunity in under four hours---a programmatic requirement designed to enable routine temporal responsiveness that is unprecedented for a space telescope of this size. Lazuli’s technical capabilities are shaped around three broad science areas: (1) time-domain and multi-messenger astronomy, (2) stars and planets, and (3) cosmology. These capabilities enable a potent mix of science spanning gravitational wave counterpart characterization, fast-evolving transients, Type Ia supernova cosmology, high-contrast exoplanet imaging, and spectroscopy of exoplanet atmospheres. While these areas guide the observatory design, Lazuli is conceived as a general-purpose facility with open time for the global community. We describe the observatory architecture and capabilities in the preliminary design phase, with science operations anticipated following a rapid development cycle from concept to launch; quoted performance specifications reflect current design requirements and modeling predictions, pending on-orbit verification.
\end{abstract}


\newacronym{AU}{AU}{astronomical Unit [1.5e11 m]}
\newacronym{pc}{pc}{parsec}
\newacronym{mas}{mas}{milliarcsecond}
\newacronym{nm}{nm}{nanometer}
\newacronym{CTE}{CTE}{coefficient of thermal expansion}
\newacronym{sqarc}{$as^2$}{square arcsecond}
\newacronym{ppm}{ppm}{Part Per Million}

% Astrophysics:
\newacronym{smc}{SMC}{Small Magellanic Cloud}
\newacronym{lmc}{LMC}{Large Magellanic Cloud}
\newacronym{ism}{ISM}{interstellar medium}
\newacronym{mw}{MW}{Milky Way}
\newacronym{epseri}{$\epsilon$ Eri}{Epsilon Eridani}
\newacronym{EKB}{EKB}{Edgeworth-Kuiper Belt}
\newacronym{CFR}{CFR}{Complete Frequency Redistribution}

% Organizations:
\newacronym{nasa}{NASA}{National Aeronautics and Space Agency}
\newacronym{esa}{ESA}{European Space Agency}
\newacronym{omi}{OMI}{$\textit{Optical Mechanics Inc.}$}
\newacronym{gsfc}{GSFC}{\gls{nasa} Goddard Space Flight Center}
\newacronym{stsci}{STScI}{Space Telescope Science Institute}
\newacronym{nsroc}{NSROC}{\gls{nasa} Sounding Rocket Operations Contract}
\newacronym{wff}{WFF}{\gls{nasa} Wallops Flight Facility}
\newacronym{wsmr}{WSMR}{White Sands Missile Range}
\newacronym{BCT}{BCT}{Blue Canyon Technologies}
\newacronym{SFL}{SFL}{Spaceflight Laboratory}
\newacronym{UA}{UA}{University of Arizona}
\newacronym{UTIAS}{UTIAS}{University of Toronto Institute for Aerospace Studies}
\newacronym{DIATF}{DIATF}{Drake Imager Assembly and Testing Facility}
\newacronym{ETS}{ets}{Engineering Technical Services}
\newacronym{ASTM}{ASTM}{American Society for Testing and Materials}

% Technologies and Sensors:
\newacronym{irac}{IRAC}{Infrared Array Camera}
\newacronym[plural=CCDs, firstplural=charge-coupled devices (CCDs)]{ccd}{CCD}{charge-coupled device}
\newacronym[plural=EMCCDs, firstplural=electron multiplying charge-coupled devices (EMCCDs)]{EMCCD}{EMCCD}{electron multiplying charge-coupled device}
\newacronym{DM}{DM}{Deformable Mirror}
\newacronym{ipc}{IPC}{Image Proportional Counter}
\newacronym{cots}{COTS}{Commercial Off-The-Shelf}
\newacronym{COTS}{COTS}{commercial off-the-shelf}
\newacronym{ISR}{ISR}{incoherent scatter radar}
\newacronym{atcamera}{AT}{angle tracker}
\newacronym{MEMS}{MEMS}{microelectromechanical systems}
\newacronym{QE}{QE}{quantum efficiency}
\newacronym{RTD}{RTD}{Resistance Temperature Detector}
\newacronym{PID}{PID}{Proportional-Integral-Derivative}
\newacronym{PRNU}{PRNU}{photo response non-uniformity}
\newacronym{DSNU}{PRNU}{dark signal non-uniformity}
\newacronym{CMOS}{CMOS}{complementary metal–oxide–semiconductor}
\newacronym{TEC}{TEC}{thermoelectric cooler}

% Optics:
\newacronym{FOV}{FOV}{field-of-view}
\newacronym{NIR}{NIR}{near-infrared}
\newacronym{PV}{PV}{Peak-to-Valley}
\newacronym{MRF}{MRF}{Magnetorheological finishing}
\newacronym{AO}{AO}{Adaptive Optics}
\newacronym{TTP}{TTP}{tip, tilt, and piston}
\newacronym{FPS}{FPS}{fine pointing system}
\newacronym{SHWFS}{SHWFS}{Shack-Hartmann Wavefront Sensor}
\newacronym{OAP}{OAP}{off-axis parabola}
\newacronym{LGS}{LGS}{laser guide star}
\newacronym{WFCS}{WFCS}{wavefront control system}
\newacronym{OPD}{OPD}{optical path difference}
\newacronym{MEL}{MEL}{Master Equipment List}
\newacronym{EFC}{EFC}{electric-field conjugation}
\newacronym{iEFC}{EFC}{implicit \gls{EFC}}
\newacronym{LDFC}{LDFC}{linear dark field control}
\newacronym{TMA}{TMA}{three-mirror anastigmat}
\newacronym{resel}{resel}{resolution element}
\newacronym{IFU}{IFU}{integral field unit}
\newacronym{IFS}{IFS}{integral field spectrograph}
\newacronym{ULE}{ULE}{ultra-low expansion glass}

% Spacecraft and Sounding Rocket:
\newacronym{acs}{ACS}{attitude control system}
\newacronym{orsa}{ORSA}{Ogive Recovery System Assembly}
\newacronym{gse}{GSE}{ground station equipment}
\newacronym{FSM}{FSM}{fast steering Mirror}
\newacronym{CFRP}{CRFP}{carbon fiber reinforced plastic}
\newacronym{CDP}{CDP}{critical design phase}
\newacronym{CDR}{CDR}{critical design review}
\newacronym{LEO}{LEO}{low-earth orbit}
\newacronym{GEO}{GEO}{geosynchronous orbit}
\newacronym{FEA}{FEA}{finite element analysis}
\newacronym{ESPA}{ESPA}{EELV Secondary Payload Adapter}
\newacronym{EEID}{EEID}{Earth-equivalent Insolation Distance, the distance from the star where the incident energy density is that of the Earth received from the Sun}
\newacronym{STOP}{STOP}{Structural-Thermal-Optical-Performance}
\newacronym{STM}{STM}{science traceability matrix}
\newacronym{ConOps}{ConOps}{concept of operations}
\newacronym{NRE}{NRE}{non-recurring engineering}
\newacronym{CSR}{CSR}{concept study report}
\newacronym{XAO}{XAO}{extreme-adaptive optics}
\newacronym{AT}{AT}{angle tracking camera}
\newacronym{SRR}{SRR}{system requirements review}
\newacronym{ROI}{ROI}{region of interest}
\newacronym{LCP}{LCP}{liquid-crystal polymer}
\newacronym{CBE}{CBE}{current best estimate}
\newacronym{SBC}{SBC}{single-board computer}
\newacronym{ADCS}{ADCS}{attitude determination and control system}
\newacronym{CDH}{C$\&$DH}{command and data handling}
\newacronym{EPS}{EPS}{electrical power system}
\newacronym{TLE}{TLE}{Two Line Element set}
\newacronym{TRL}{TRL}{technology readiness level}
\newacronym{swap}{SWaP}{Size, Weight, and Power}

% Electronics / Radiation
\newacronym{DAC}{DAC}{digital-to-analog converter}
\newacronym{SEE}{SEE}{Single Event Effects}
\newacronym{MOSFET}{MOSFET}{Metal-Oxide-Semiconductor Field-Effect Transistor}
\newacronym{TID}{TID}{Total Ionizing Dose}
\newacronym{TNID}{TNID}{Total Non-Ionizing Dose}
\newacronym{EDAC}{EDAC}{Error Detection and Correction}
\newacronym{EMI}{EMI}{Electromagnetic Interference}
\newacronym{EMC}{EMC}{Electromagnetic Compatibility}

% High Contrast Imaging:
\newacronym{WFS}{WFS}{wavefront sensor}
\newacronym{LSI}{LSI}{Lateral Shearing Interferometer}
\newacronym{VVC}{VVC}{Vector Vortex Coronagraph}
\newacronym{VNC}{VNC}{Visible Nulling Coronagraph}
\newacronym{CGI}{CGI}{Coronagraph Instrument}
\newacronym{IWA}{IWA}{Inner Working Angle}
\newacronym{OWA}{OWA}{Outer Working Angle}
\newacronym{NPZT}{N-PZT}{Nuller piezoelectric transducer}
\newacronym{ZWFS}{ZWFS}{Zernike wavefront sensor}
\newacronym{SPC}{SPC}{Shaped Pupil Coronagraph}
\newacronym{HLC}{HLC}{Hybrid-Lyot Coronagraph}
\newacronym{ADI}{ADI}{angular differential imaging}
\newacronym{RDI}{RDI}{reference differential imaging}
\newacronym{HOWFSC}{HOWFS/C}{high-order wavefront sensing and control}
\newacronym{WFSC}{WFSC}{wavefront sensing and control}
\newacronym{CDI}{CDI}{coherent differential imaging}
\newacronym{SCC}{SCC}{self-coherent camera}
\newacronym{LLOWFS}{LLOWFS}{Lyot low-order wavefront sensor}
\newacronym{VSG}{VSG}{vacuum surface gauge}
\newacronym{TDEM}{TDEM}{Technology Development for Exoplanet Missions}
\newacronym{HZ}{HZ}{habitable zone}
\newacronym{ODI}{ODI}{orbital-differential imaging}
\newacronym{PSFTFC}{PSFTFC}{PSF template subtracted coronagraphy}
\newacronym{LOWFSC}{LOWFSC}{Low-order \gls{WFS} and control}
\newacronym{scoob}{SCoOB}{Space Coronagraph Optical Bench}
\newacronym{FDPR}{FDPR}{focus diversity phase retrieval}

% Observatories and Instruments:
\newacronym{HST}{HST}{Hubble Space Telescope}
\newacronym{GPS}{GPS}{Global Positioning System}
\newacronym{ISS}{ISS}{International Space Station}
\newacronym[description=Advanced CCD Imaging Spectrometer]{acis}{ACIS}{Advanced \gls{ccd} Imaging Spectrometer}
\newacronym{stis}{STIS}{$\textit{Space Telescope Imaging Spectrograph}$}
\newacronym{mcp}{MCP}{Microchannel Plate}
\newacronym{jwst}{JWST}{$\textit{JWST}$}
\newacronym{fuse}{FUSE}{$\textit{FUSE}$}
\newacronym{galex}{GALEX}{$\textit{Galaxy Evolution Explorer}$}
\newacronym{spitzer}{Spitzer}{$\textit{Spitzer Space Telescope}$}
\newacronym{mips}{MIPS}{Multiband Imaging Photometer for \gls{spitzer}}
\newacronym{gissmo}{GISSMO}{Gas Ionization Solar Spectral Monitor}
\newacronym{iue}{IUE}{International Ultraviolet Explorer}
\newacronym{spinr}{SPINR}{$\textit{Spectrograph for Photometric Imaging with Numeric Reconstruction}$}
\newacronym{imager}{IMAGER}{$\textit{Interstellar Medium Absorption Gradient Experiment Rocket}$}
\newacronym{TPF-C}{TPF-C}{Terrestrial Planet Finder Coronagraph}
\newacronym{RAIDS}{RAIDS}{Atmospheric and Ionospheric Detection System }
\newacronym{mama}{MAMA}{Multi-Anode Microchannel Array}
\newacronym{ATLAST}{ATLAST}{Advanced Technology Large Aperture Space Telescope}
\newacronym{PICTURE}{PICTURE}{Planet Imaging Concept Testbed Using a Rocket Experiment}
\newacronym{LITES}{LITES}{Limb-imaging Ionospheric and Thermospheric Extreme-ultraviolet Spectrograph}
\newacronym{LBT}{LBT}{Large Binocular Telescope}
\newacronym{LBTI}{LBTI}{Large Binocular Telescope Interferometer}
\newacronym{KIN}{KIN}{Keck Interferometer Nuller}
\newacronym{SHARPI}{SHARPI}{Solar High-Angular Resolution Photometric Imager}
\newacronym{IRAS}{IRAS}{Infrared Astronomical Satellite}
\newacronym{HARPS}{HARPS}{High Accuracy Radial velocity Planetary}
\newacronym{hstSTIS}{STIS}{Space Telescope Imaging Spectrograph}
\newacronym{spitzerIRAC}{IRAC}{Infrared Array Camera}
\newacronym{spitzerMIPS}{MIPS}{Multiband Imaging Photometer for Spitzer}
\newacronym{spitzerIRS}{IRS}{Infrared Spectrograph}
\newacronym{CHARA}{CHARA}{Center for High Angular Resolution Astronomy}
\newacronym{wfirst-afta}{WFIRST-AFTA}{Wide-Field InfrarRed Survey Telescope-Astrophysics Focused Telescope Assets}
\newacronym{GPI}{GPI}{Gemini Planet Imager}
\newacronym{WFIRST}{Roman}{Nancy Grace Roman Space Telescope}
\newacronym{HabEx}{HabEx}{Habitable Exoplanet Observatory Mission Concept}
\newacronym{FGS}{FGS}{Fine Guidance Sensor}
\newacronym{MGHPCC}{MGHPCC}{Massachusetts Green High Performance Computing Center}
\newacronym{WISE}{WISE}{Wide-field Infrared Survey Explorer}
\newacronym{ALMA}{ALMA}{Atacama Large Millimeter Array}
\newacronym{GRAIL}{GRAIL}{Gravity Recovery and Interior Laboratory}
\newacronym{jwstNIRCam}{NIRCam}{near-\gls{IR}-camera}
\newacronym{jwstMIRI}{MIRI}{Mid-Infrared Instrument}
\newacronym{LUVOIR}{LUVOIR}{Large UV Optical IR Surveyor}
\newacronym{Roman}{Roman}{Nancy Grace Roman Space Telescope}
\newacronym{STP}{STP}{Space Telescope Pathfinder}
\newacronym{UM}{UM}{Ultramarine}
\newacronym{HLST}{HLST}{Hypothetical Large Space Telescopes}
\newacronym{CDEEP}{ESC}{STP ExtraSolar Camera, a Coronagraphic Pathfinder}
\newacronym{ESC}{ESC}{ExtraSolar Camera}
\newacronym{CCS}{WCC}{Wavefront and Context Camera}
\newacronym{WCC}{WCC}{Wavefront and Context Camera}
\newacronym{LSST}{LSST}{Large Synoptic Survey Telescope}
\newacronym{M1}{M1}{Mirror 1 (Telescope Module Primary Mirror)}
\newacronym{M2}{M2}{Mirror 2 (Telescope Module Secondary Mirror)}
\newacronym{M3}{M3}{Mirror 3}
\newacronym{M4}{M4}{Mirror 4}
\newacronym{AOA}{AOA}{Aft-Optics Assembly}
\newacronym{FOA}{FOA}{Fore-Optics Assembly}
\newacronym{AOSS}{AOSS}{Aft-optics Support Structure}
\newacronym{MMTO}{MMTO}{MMT Observatory}
\newacronym{MMT}{MMT}{Multiple Mirror Telescope}
\newacronym{PMCC}{PMCC}{primary mirror (M1) control computer}
\newacronym{PMSS}{PMSS}{primary mirror (M1) support structure}
\newacronym{UASAL}{UASAL}{UArizona Space Astrophysics Lab}
\newacronym{ITL}{ITL}{Imaging Technology Lab}
\newacronym{TAO}{TAO}{Tokyo Atacama Observatory}
\newacronym{UVS}{UVS}{Ultraviolet Spectrograph}
\newacronym{STIS}{STIS}{Space Telescope Imaging Spectrograph}
\newacronym{SCoOB}{scoob}{space-coronagraph optical bench}
\newacronym{HWO}{HWO}{Habitable Worlds Observatory}

% Software:
\newacronym{AURIC}{AURIC}{The Atmospheric Ultraviolet Radiance Integrated Code}
\newacronym{FFT}{FFT}{Fast Fourier Transform  }
\newacronym{MODTRAN}{MODTRAN   }{ MODerate resolution atmospheric TRANsmission }
\newacronym{idl}{IDL}{$\textit{Interactive Data Language}$}
\newacronym[sort=NED,description=NASA/IPAC Extragalactic Database]{ned}{NED}{\gls{nasa}/\gls{ipac} Extragalactic Database}
\newacronym{iraf}{IRAF}{Image Reduction and Analysis Facility}
\newacronym{wcs}{WCS}{World Coordinate System}
\newacronym{pegase}{PEGASE}{$\textit{Projet d'Etude des GAlaxies par Synthese Evolutive}$}
\newacronym{dirty}{DIRTY}{$\textit{DustI Radiative Transfer, Yeah!}$}
\newacronym{CUDA}{CUDA}{Compute Unified Device Architecture}
\newacronym{KLIP}{KLIP}{Karhunen-Lo`eve Image Processing}
\newacronym{FEM}{FEM}{finite element method}
\newacronym{CLI}{CLI}{command-line interface}
\newacronym{CAD}{CAD}{computer-aided design}
\newacronym{DBMS}{DBMS}{database management system}
\newacronym{POPPY}{POPPY}{Physical Optics Propagation in Python}
\newacronym{SOEDMS}{SOEDMS}{Steward Observatory Electronic Data Management System}
\newacronym{CAAO}{CAAO}{Center for Astronomical Adaptive Optics}

% Earth's Atmosphere and Ionosphere:
\newacronym{MSIS}{MSIS}{Mass Spectrometer Incoherent Scatter Radar}
\newacronym{nmf2}{$N_m$}{F2-Region Peak density}
\newacronym{hmf2}{$h_m$}{F2-Region Peak height}
\newacronym{H}{$H$}{F2-Region Scale Height}

% Misc jargon:
\newacronym{isr}{ISR}{Incoherent Scatter Radar}
\newacronym[description=TLA Within Another Acronym]{twaa}{TWAA}{\gls{tla} Within Another Acronym}
\newacronym[plural=SNe, firstplural=Supernovae (SNe)]{sn}{SN}{Supernova}
\newacronym{EUV}{EUV}{Extreme-Ultraviolet}
\newacronym{EUVS}{EUVS}{\gls{EUV} Spectrograph}
\newacronym{F2}{F2}{Ionospheric Chapman F Layer}
\newacronym{F10.7}{F10.7}{ 10.7 cm radio flux [10$^{-22}$ W m$^{-2}$ Hz$^{-1}$]  }
\newacronym{FUV}{FUV}{far-ultraviolet}
\newacronym{IR}{IR}{infrared}
\newacronym{MUV}{MUV}{mid-ultraviolet }
\newacronym{NUV}{NUV}{near-ultraviolet}
\newacronym{nir}{NIR}{near-infrared}
\newacronym{mir}{MIR}{mid-infrared}
\newacronym{UV}{UV}{ultraviolet}
\newacronym{O$^+$}{O$^+$}{Singly Ionized Oxygen  Atom}
\newacronym{OI}{OI}{Neutral Atomic Oxygen Spectroscopic State}
\newacronym{OII}{OII}{Singly Ionized Atomic Oxygen Spectroscopic State}
\newacronym{PSF}{PSF}{point spread function}
\newacronym{$R_E$}{$R_E$}{Earth radii [$\approx$ 6400 km] }
\newacronym{RV}{RV}{radial velocity}
\newacronym{WFE}{WFE}{wavefront error}
\newacronym{sed}{SED}{spectral energy distribution}
\newacronym[plural=PAHs, firstplural=Polycyclic Aromatic Hydrocarbons (PAHs)]{pah}{PAH}{Polycyclic Aromatic Hydrocarbon}
\newacronym{obsid}{OBSID}{Observation Identification}
\newacronym{SZA}{SZA}{Solar Zenith Angle}
\newacronym{PZT}{PZT}{lead zirconate titanate}
\newacronym{AIT}{AIT}{Assembly, Integration and Test}
\newacronym{CPR}{CPR}{cost performance report}
\newacronym{FDR}{FDR}{final design review}
\newacronym{DHS}{DHS}{data handling system}
\newacronym{CNS}{CNS}{communications and network system}
\newacronym{FOC}{FOC}{fiber optic cable}
\newacronym{CDS}{CDS}{correlated double sampling}
\newacronym{DDCP}{DDCP}{document and drawing control plan}
\newacronym{FOCS}{FOCS}{feed optics control system}
\newacronym{CSH}{CSH}{camera systems hardware}
\newacronym{CSS}{CSS}{camera systems software}
\newacronym{DMA}{DMA}{dynamic mechincal anlysis}
\newacronym{MWIR}{MWIR}{midwave infrared}
\newacronym{LWIR}{LWIR}{longwave infrared}
\newacronym{SVD}{SVD}{singular value decomposition}
\newacronym{NRM}{NRM}{normal response mode}
\newacronym{KISS}{KISS}{keep it sans spectrometer}
\newacronym{CIDL}{CIDL}{configuration item data list}
\newacronym{ICD}{ICD}{interface control document}
\newacronym{ERD}{ERD}{Environmental Requirements Document}
\newacronym{EP}{EP}{Telescope Entrance Pupil}
\newacronym{HFOV}{HFOV}{Half Field of View}
\newacronym{L3}{L3}{Telescope Module Collimating Lens}
\newacronym{QKD}{QKD}{Quantum Key Distribution}
\newacronym{XP}{XP}{Telescope Exit Pupil}
\newacronym{TBC}{TBC}{To Be Confirmed}
\newacronym{TBD}{TBD}{To Be Determined}
\newacronym{TBR}{TBR}{To Be Reviewed}
\newacronym{EEIS}{EEIS}{End-to-End Information Systems}
\newacronym{EA}{EA}{Executing Agent}
\newacronym{SPGD}{SPGD}{Stochastic Parallel Gradient Descent}
\newacronym{TAM}{TAM}{Test Allocation Matrix}
\newacronym{CCP}{CCP}{Contamination Control Plan}
\newacronym{VC}{VC}{Visualy Clean}
\newacronym{VC-S}{VC-S}{Visualy Clean Sensitive}
\newacronym{VC-HS}{VC-HS}{Visualy Clean High Sensitive}
\newacronym{PAC}{PAC}{Percent Area Coverage}
\newacronym{ATLO}{ATLO}{Assembly Test, and Launch Operations}
\newacronym{QA}{QA}{Quality Assurance}
\newacronym{UUT}{UUT}{Unit Under Test}
\newacronym{P/N}{P/N}{Part Number}
\newacronym{ESD}{ESD}{Electro-static Discharge}
\newacronym{AC}{AC}{Alternating Current}
\newacronym{RH}{RH}{Relative Humidity}
\newacronym{RGA}{RGA}{Residual Gas Analyzer}
\newacronym{esds}{ESDS}{ESD Sensitive}
\newacronym{dmm}{DMM}{Digital Multimeter}
\newacronym{DC}{DC}{Direct Current}
\newacronym{CPG}{CPG}{Common Point Ground}
\newacronym{WM}{WM}{Workmanship Manual}
\newacronym{N/A}{N/A}{Not Applicable}
\newacronym{na}{NA}{Not Applicable}
\newacronym{MM}{MM}{Machine Model [for electrostatic discharge]}
\newacronym{LVDS}{LVDS}{Low-Voltage Differential Signal}
\newacronym{LNA}{LNA}{Low Noise Amplifier}
\newacronym{FoS}{FoS}{Factors of Safety}
\newacronym{DLL}{DLL}{Design Limit Loads}
\newacronym{MoS}{MoS}{Margin of Safety = (Material Allowable / (Max Stress \gls{MPE} * \gls{FoS})) - 1}
\newacronym{MPE}{MPE}{Maximum Predicted Environments}

% Material Abbreviations
\newacronym{PVC}{PVC}{Polyvinyl Chloride}
\newacronym{PTFE}{PTFE}{Polytetrafluoroethylene (Teflon)}

% Statistics:
\newacronym{PCA}{PCA}{principal component analysis}
\newacronym{fwhm}{FWHM}{full-width-half maximum}
\newacronym{RMS}{RMS}{root mean squared}
\newacronym{RMSE}{RMSE}{root mean squared error}
\newacronym{MCMC}{MCMC}{Marcov chain Monte Carlo}
\newacronym{DIT}{DIT}{discrete inverse theory}
\newacronym{SNR}{SNR}{signal-to-noise ratio}
\newacronym{PSD}{PSD}{power spectral density}
\newacronym{NMF}{NMF}{non-negative matrix factorization}
\newacronym{DOF}{DOF}{degrees-of-freedom}

% Management
\newacronym{BOM}{BOM}{Bill of Materials}
\newacronym{POC}{POC}{point of contact}
\newacronym{CDRL}{CDRL}{Contract Data Requirement List}
\newacronym{FBD}{FBD}{Functional Block Diagram}
\newacronym{ESC}{ESC}{ExtraSolar Coronagraph}
\newacronym{WCC}{WCC}{Widefield Context Camera}
\newacronym{LP}{LP}{linear polarizer}
\newacronym{QWP}{QWP}{quarter-wave plate}
\newacronym{HOWFS}{HOWFS}{high-order wavefront sensing}
\newacronym{VVW}{VVW}{vector-vortex waveplate}
\newacronym{ELT}{ELT}{extremely large telescope}

% ----------------------------------------------
% ----------------------------------------------
% ----------------------------------------------
\section{A Philanthropic Approach to Astrophysics} 

Over the past decade, the landscape of space-based astronomy has shifted toward greater emphasis on rapid development, focused instrument suites, and responsiveness to time-critical science. The Lazuli Space Observatory is designed to embody this shift, with two primary goals: to deploy and operate a world-class astronomical observatory in space, and to do so on a substantially accelerated and lower-cost development timeline relative to traditional approaches. The project is underway, with secured funding and a defined budget, as well as detailed plans for the instruments, spacecraft, and development schedule.

The Lazuli Space Observatory is part of a larger program---the Eric and Wendy Schmidt Observatory System---which will include at least three ground-based observatories as well as one or more space-based observatories. Each ground-based telescope adopts a similarly risk-tolerant approach, leveraging large numbers of smaller components and modern computing to achieve scalable performance. All of these facilities are pure research instruments intended to enable deeper understanding of the Universe. They are designed to support global community use through rapid, open dissemination of data.

The Lazuli Space Observatory concept presented here draws from an initial mission concept~\citep{Perlmutter2020_QuickToLaunch,Perlmutter2021_STP101} and a subsequent feasibility study led through the Space Sciences Laboratory at University of California, Berkeley, which articulated a high-risk, high-speed paradigm for large-aperture space astronomy on large-capacity launch vehicles, exemplified by a first mission combining an integral-field spectrograph and an imager to address timely dark-energy science, as well as the nimble follow-up of gravitational-wave events, exoplanet transits, and other transients. The feasibility study included a coronagraph that was developed by \citet{douglas_approaches_2023}, with active wavefront control based on a laboratory-tested design. We initially explored various designs, prototypes, and evaluations for this notional 6.5m mission focused on much more specific science objectives. That approach ultimately proved infeasible within our desired time, risk, and financial constraints.

In late 2024, the project pivoted to the observatory architecture described here, which was approved for construction in mid-2025. The Lazuli Space Observatory employs a 3m primary mirror and a broader, more detailed science program designed to take full advantage of the coronagraph, camera, and spectrograph.

The Lazuli Space Observatory is funded privately, a first for a space mission of this scale. Philanthropic funding can help fill the gap between relatively small, rapid missions and very ambitious but expensive and decades-long flagship projects.
Engineering for the Lazuli Space Observatory is based on existing cutting edge technologies, while project management builds on decades of experience across government, academic, and industrial space programs, incorporating lessons learned from the development and operations of large space telescope facilities, as well as the broader commercial space sector. The instrument suite will enable a wide range of important observing campaigns while remaining deliberately constrained to limit observatory complexity. The Lazuli Space Observatory is designed to be launchable within approximately 3--5 years of the start of detailed planning. This effort is intended to demonstrate a viable pathway for deploying significant astronomical instrumentation on accelerated timescales. 

This paper is the first in a series describing the Lazuli Space Observatory and its scientific capabilities. It begins by outlining the science motivation for Lazuli (\S \ref{sec:sciencemotivation}) and the broader development approach it represents, followed by a description of the mission design guidelines that translate those motivations into concrete architectural choices  (\S \ref{sec:missiondesign}). We then present the observatory architecture and design in detail (\S \ref{sec:mission}), followed by the scientific capabilities enabled by those designs (\S \ref{sec:science}). \S \ref{sec:operations} describes the mission operations concept, including orbit selection and scheduling strategies optimized for rapid response. \S \ref{sec:community} outlines community access, data policies, software, and engagement frameworks. We conclude in \S \ref{sec:conclusion} with a synthesis of Lazuli’s role within the evolving landscape of space-based astrophysics.

% ----------------------------------------------
\section{Science Motivation}
\label{sec:sciencemotivation}
Lazuli is designed to move cutting-edge technology into the hands of astronomers faster than traditional mission development timelines, with a goal to accelerate exciting astrophysical discoveries. This approach accepts a higher risk profile in exchange for rapid scientific return, and provides the opportunity to fly promising but lower-heritage technologies that might otherwise await slow derisking timelines. Our science capabilities are chosen to be both transformative for this decade and achievable within a 3–-5 year development timeline from concept to launch. Even though the Lazuli team approached this question from the perspective of rapid development for impactful science, our priorities are well aligned with the recommendations of the Astro2020 Decadal Survey \citep{Decadal2021} and complement existing and planned observatories by performing precursor observations allowing target and technique optimizations for future missions, technology maturation, and follow-up of high value targets.

\subsection{Rapidly Responding to a Transient Universe}
The advent of wide-field time-domain surveys has revealed a transient universe that demands rapid spectroscopic follow-up. Gravitational wave electromagnetic counterparts fade rapidly, often within hours to days \citep[e.g., GW170817;][]{Abbott2017}; Fast Blue Optical Transients (FBOTs) evolve on hour-to-day timescales \citep[e.g., AT2018cow;][]{Prentice2018}; and early-time observations of supernovae provide critical constraints on progenitor systems and explosion physics \citep[e.g.,][]{2017hsn..book..967W}. Current large space-based facilities, while exquisitely sensitive, respond to targets of opportunity on timescales of days to weeks---often too late to capture the most rapidly-evolving phenomena. A notable exception is Swift, which can respond to ToOs on timescales of minutes but has a much smaller aperture. Ground-based telescopes can respond quickly but face fundamental limitations: atmospheric emission and absorption, weather and diurnal interruptions, and seeing-limited resolution. With large sky surveys from the Rubin Observatory \citep{Ivezic2019}, Nancy Grace Roman Space Telescope \citep{Spergel2015}, Argus \citep{Law2022} and ULTRASAT \citep{2024ApJ...964...74S} imminent, and upgraded gravitational wave detectors coming online in the late 2020s, the astronomical community will discover transients at unprecedented rates—but lacks a space-based platform to characterize them spectroscopically within hours. Lazuli addresses this gap directly. With a threshold response time $<4$ hours from trigger receipt to first observation (with a goal of 90 minutes), the observatory is designed to capture transient phenomena during their earliest evolutionary phases—when physical conditions change most rapidly.

\subsection{Flying Ambitious Technology to Accelerate Science Readiness}
Several of Lazuli’s key capabilities rely on technologies with limited space flight heritage but strong technical maturity and high scientific potential. 
For example, high-contrast imaging of exoplanets from space offers exceptional potential for understanding planetary systems, but the technique requires iteration between laboratory demonstrations and on-orbit experience to achieve the contrasts needed for detecting true Earth analogs \citep{Ruane2018, Kasdin2020,mennesson_current_2024}. Lazuli embraces the opportunity to fly coronagraphic technologies with low space heritage but high science potential for science usage now---accepting higher risk in exchange for expanded astrophysical returns and operational lessons that cannot be learned on the ground \citep{douglas_approaches_2023}. The goal of flying such technology on Lazuli is to produce valuable science in its own right while building the heritage needed for future flagship missions such as the Habitable Worlds Observatory \citep{Decadal2021}. This approach---prototyping technology and proceeding directly to flight rather than waiting for hierarchical derisking---accelerates both scientific return and technical readiness for the missions that will ultimately perform deeper searches for life beyond Earth.

\subsection{Sustaining Large-Aperture Optical-NIR Capability in Space}
Finally, Lazuli addresses a growing need for large-aperture optical-NIR capability in space. The Hubble Space Telescope (HST) has served the astronomical community for over three decades and remains scientifically productive, but as the sole flagship-class space observatory operating at optical and ultraviolet wavelengths it faces intense proposal pressure \citep{HST_status}, and no planned facility provides equivalent or stronger capabilities until the Habitable Worlds Observatory in the early 2040s \citep{Decadal2021}.
The James Webb Space Telescope (JWST), while transformative in the infrared, operates at wavelengths redder than 600~nm \citep{Gardner2006}, and is already facing the highest proposal pressure of any large observatory \citep{Rao2024JWSTDemand}. The upcoming Roman Space Telescope, scheduled for launch in 2026, will image the near-IR sky at wavelengths redder than 500~nm and will acquire $R \sim 100-600$ spectroscopy at $\lambda > 750$ nm \citep{Spergel2015, Akeson2019Roman} but is optimized for wide-field survey science rather than rapid-response, targeted spectroscopic follow-up or high-cadence observations. Lazuli instead occupies a complementary observational niche. It is designed to provide rapid-response, diffraction-limited optical and near-infrared imaging and spectroscopy, with an operations model centered on time-critical observations and instruments that incorporate decades of advances in optical/NIR detector technology, instrument design, and mission operations. These capabilities differ from, rather than replace, HST’s strengths. In the spirit of HST's enduring legacy \citep[e.g.][]{roman1974LST}, Lazuli aims to serve a broad range of existing science needs while holding capability for ideas yet to come. 

% ----------------------------------------------
\section{Mission Design Guidelines}
\label{sec:missiondesign}
Lazuli represents an experiment in space observatory development for astrophysics, borrowing from successful `new space' approaches in industry. Rather than following the traditional flagship model---where technology is invented, matured over decades, and flown for science use after comprehensive risk reduction---Lazuli aims to demonstrate a different development curve. Rather than envision and create new technology ab initio, we apply available technology in novel ways---building on the heritage of research and development from major international partnerships and decades of ground-based and space-based astronomy.

The mission design is guided by a set of Level-0 (L0) program objectives that flow from two overarching goals: to demonstrate a different and more rapid approach to large-aperture space observatory development, and to provide a world-class astrophysics facility to the global scientific community. The following principles shape major design decisions for Lazuli:

\begin{itemize}
    \item \textbf{Schedule as a feature not a constraint.} Lazuli is designed to launch and operate while the science questions it addresses remain pressing and while synergistic facilities---including Rubin Observatory, the Roman Space Telescope, and gravitational wave detector networks---are active. A compressed development timeline is not a compromise but a deliberate design choice that maximizes scientific relevance and responsiveness to community needs.

    \item \textbf{Risk tolerance as a driver of cost discipline.} Lazuli is intentionally positioned in a different region of the cost–risk trade space than traditional flagship missions, enabling accelerated development and near-term scientific return. This posture constrains cost by reducing the need for prolonged technology maturation and exhaustive pre-flight risk retirement, with risk acceptance bounded through focused requirements, selective use of heritage, and an operations concept designed for on-orbit learning. Unlike some flagship observatories and designs, Lazuli is not designed to be serviceable, keeping cost and complexity down.
    
    \item \textbf{Rapid response as a primary design driver.} The capability to observe transient phenomena within hours of discovery is not an enhancement but rather a foundational requirement that shapes spacecraft design, mission operations, and ground system architecture.

    \item \textbf{Focused instrument suite.} Limiting the number of modes, as well as focusing on core performance verification, mitigates the schedule delays and cost growth that commonly arise when complex instrument suites with competing requirements must achieve simultaneous readiness.
    
    \item \textbf{Coordination over competition.} Lazuli is intended to complement and enhance the scientific return of other facilities rather than replicate their capabilities. Data will be released promptly to maximize benefit to the broader community.

    \item \textbf{General-purpose by intent.} Specific science cases serve as tools to anchor broad capability, but Lazuli is conceived as a general-purpose facility—capable of supporting research well beyond what is described here, including questions that have not yet been formulated---maintaining flexibility for investigations that will emerge over the mission lifetime.
 
\end{itemize}

% ----------------
\begin{figure*}[t!]
\centering
\includegraphics[width=0.99\textwidth]{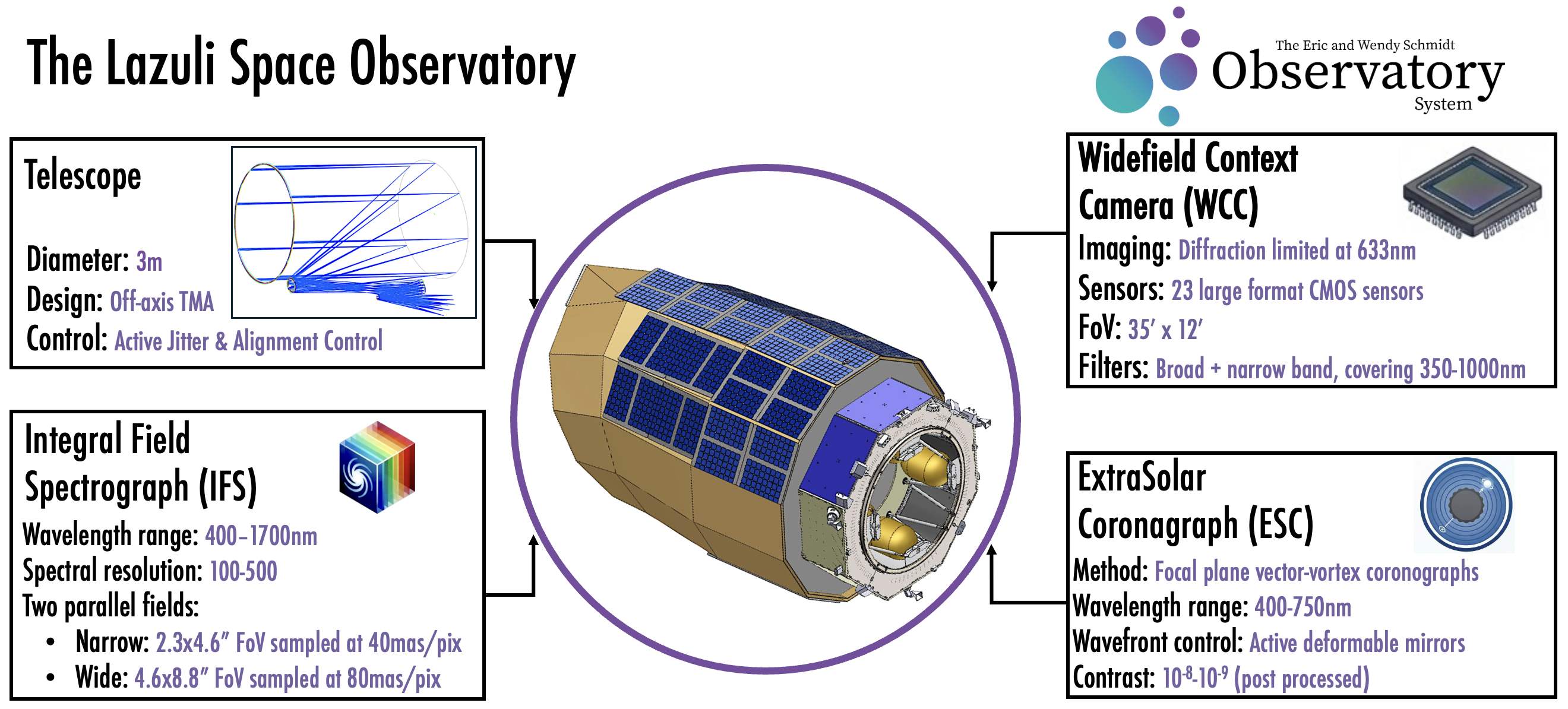}
\caption{Overview of the 3~m Lazuli Space Observatory and its three instruments, the Widefield Context Camera (WCC), the Integral Field Spectrograph (IFS), and the ExtraSolar Coronagraph (ESC). Main properties and characteristics of the telescope and the instruments are highlighted.}
\label{fig:overview}
\end{figure*}

These principles inform the observatory architecture, operations concept, and science capabilities described in the sections that follow. Lazuli's development has been driven by an iterative process: technical capabilities were chosen to enable specific science goals, while the science scope has evolved in response to realistically achievable technical performance---a parallel development approach that reflects the mission's compressed timeline, a strategy consistent with agile systems engineering best practices. 

% ----------------------------------------------
\section{Observatory Architecture \& Design Capabilities}\label{sec:mission}
\subsection{Observatory Overview}
The Lazuli Space Observatory will be a dedicated 4,000~kg space-based astronomical facility that will operate in a highly elliptical lunar-resonant orbit. The observatory comprises an off-axis three-mirror anastigmat (TMA) telescope with a 3-meter primary mirror unobscured by the secondary mirror---a configuration that delivers diffraction-limited Point Spread Function (PSF) quality across a wide focal plane (\S~\ref{sec:ota}). The observatory has a nominal lifetime requirement of three years and is provisioned with consumables to support a 10-year extended-mission goal.

Three science instruments provide complementary capabilities spanning imaging, spectroscopy, and high-contrast coronagraphy. The Widefield Context Camera (WCC; \S~\ref{sec:wcc}) delivers imaging across 350--1000~nm with broad-band and narrow-band filters, in-focus and defocused configurations, and high-cadence readout modes. The Integral Field Spectrograph (IFS; \S~\ref{sec:ifs}) provides continuous spectral coverage from 400--1700~nm at a spectral resolution $R \sim 100-500$. The ExtraSolar Coronagraph (ESC; \S~\ref{sec:esc}) employs a vector-vortex coronagraph with active wavefront control, leveraging the effectively unobscured aperture to enable high-contrast imaging of nearby extrasolar systems. Figure~\ref{fig:overview} provides an overview of the observatory and its instruments; Table~\ref{tab:parameters} summarizes key parameters.

\begin{deluxetable*}{ll}
\tablecaption{Overview of the Lazuli Space Observatory, the telescope, instruments, and key parameters. Throughputs and detector noise properties are listed as modeled beginning of life specifications. \label{tab:parameters}}
\tabletypesize{\scriptsize}
\tablehead{\colhead{~~~Parameter}                                 &  \colhead{Value}                        }
\startdata
\multicolumn{2}{l}{\hspace{-0.3cm} \textbf{Telescope:}}  \\
Optical layout                        & Off-axis Three-Mirror Anastigmat (TMA) \\
Primary mirror                        & 3.06~m (effective diameter)\\
F/\# and FoV                          & F/15, $0.5\degree\times0.25\degree$\\
Image Quality (target)                & Strehl ratio $>$0.8 at 633~nm (incl. jitter) \\ \hline
\multicolumn{2}{l}{\hspace{-0.3cm} \textbf{Widefield Context Camera (WCC):}}  \\ 
Wavelength Range                      & 350--1000~nm\\
Instrument + Telescope Throughput     & $>$50\% at 600~nm \\
Science Sensors ($\times15$)          & \textbf{Model \& Packaging:} Sony IMX 455\\
                                      & \textbf{Sensor size:} $9568\times6380$ pixels, $864\mathrm{mm^2}$\\
                                      & \textbf{Plate scale:} 17mas/pix for each $3.76 \mathrm{\mu m}$ pixel \\
                                      & \textbf{Read noise:} $<2e^{-}$\\
                                      & \textbf{Dark noise:} 0.0015 e/s/pix (current best estimate at $-20$C) \\
Science + Guide Sensors ($\times8$)$^\dagger$ & \textbf{Model \& Packaging:} BAE qCMOS HWK 4123  \\
                                      & \textbf{Sensor size:} $4096\times 2304$ pixels, $200\mathrm{mm^2}$\\
                                      & \textbf{Plate scale:} 21mas/pix for each $4.6 \mathrm{\mu m}$ pixel \\
                                      & \textbf{Read noise:} $<0.3e^{-}$\\
                                      & \textbf{Dark noise:} 0.004 e/s/pix (current best estimate at $-20$C) \\ \hline
\multicolumn{2}{l}{\hspace{-0.3cm} \textbf{Integral Field Spectrograph (IFS):}}  \\
Wavelength Range                      & 400--1700~nm\\
Instrument Throughput                 & Threshold: $>$40\% from 400--1000nm; $>$50\% from 1000--1700~nm \\
Detector                              & Teledyne H4RG-10 HgCdTe (1700~nm cutoff) \\
Spectral resolution$^\ddagger$        & 100--500 \\
Observing Fields                      & \textbf{Narrow-Field:} $2.3 \times 4.6\arcsec$ FOV sampled at $40$~mas/pix\\
                                      & \textbf{Wide-Field:} $4.6 \times 8.8\arcsec$ FOV sampled at $80$~mas/pix \\ \hline
\multicolumn{2}{l}{\hspace{-0.3cm} \textbf{ExtraSolar Coronagraph (ESC):}}  \\
Wavelength Range                      & \textbf{Blue Arm:} 400--540~nm\\ & \textbf{Red Arm:} 560--750~nm with multiple ($>$5) filters\\
Instrument Throughput                 & $\geq$2\% at 630~nm \\
Wavefront Control                     & Active deformable mirrors \\
Inner (IWA) and outer working angles (OWA)          & IWA$\leq$0.15\arcsec~(goal of 0.12\arcsec); OWA$\geq$0.4\arcsec~(goal of 0.6\arcsec) at 630~nm \\
(Anticipated) Raw Contrast                          & $10^{-8}$ \\
(Anticipated) Post Processed Contrast               & $10^{-9}$ \\ \hline
\multicolumn{2}{l}{\hspace{-0.3cm} \textbf{Orbit:}}  \\
Type                                  & 3:1 lunar resonant orbit\\
Period                                & 9 days\\
Perigee \& Apogee                     & 70,000 km, 285,000 km \\
Min field coverage                    & 130 days \\ 
Field of regard                       & 24,200 deg$^2$ (2.35$\pi$ sr.; goal 3.3$\pi$ sr.)\\
Continuous Viewing Zone               & Ecliptic latitude $| \beta | \ge 54 \degree$\\
\enddata
\tablenotetext{\dagger}{Although during regular operations, one qCMOS sensor is envisioned to be used for active guiding, the qCMOS sensors will also be available for scientific observations, especially to capitalize on their low read noise and sensitivity at redder wavelengths.}
\tablenotetext{\ddagger}{Using a prism dispersing element the spectral resolution reaches a minimum of $\sim$100 at 1000~nm and rises to $\sim$500 at the blue, and to $\sim$200 at the red ends, respectively.}
\end{deluxetable*}

% ----------------------------------------------
\subsection{Spacecraft Bus}\label{sec:spacecraftandbus}
The science payload will be integrated with a flight-proven spacecraft bus that provides propulsion, attitude and orbit control, power, communications, and command and data handling, while maintaining independent thermal management. The bus design leverages heritage avionics and propulsion architectures derived from prior space missions, including lunar-resonant and deep-space platforms, while employing a bespoke structure optimized for high pointing stability. The payload optical bench utilizes low coefficient of thermal expansion (CTE) materials, while body-mounted solar panels eliminate appendage modes that could disturb pointing stability.

Three-axis attitude determination is provided by three star trackers and two inertial reference units based on solid-state Coriolis vibratory gyroscopes. Six reaction wheels in a skewed configuration provide torque authority for slewing, attitude maintenance, and momentum management. At the payload level, residual line-of-sight motion is measured using guide-star observations from the WCC and corrected by the telescope Fast Steering Mirror.

The observatory is designed for compatibility with multiple launch vehicles (LVs) via a standard 2.6-meter interface, with injection to a super-synchronous transfer orbit. The monopropellant hydrazine propulsion system provides $\delta v \sim$450~m s$^{-1}$, of which $\sim$250~m s$^{-1}$ is allocated to the transfer to the operational Highly Elliptical Orbit (HEO) via a lunar gravity assist. The remaining propellant is allocated to attitude momentum management, contingency, and margin.  The lunar-resonant orbit operates outside Earth's trapped radiation belts, reducing radiation-induced noise and total ionizing dose environment, while providing thermal stability for cryogenic systems (\S~\ref{sec:orbit}). Continuous access to a commercial ground station network enables high-bandwidth science data downlink and uplink commanding for real-time observation tasking. The system is designed to deliver an average of 70 GB day$^{-1}$ of mission data. Line-of-Sight (LOS) stability is achieved through a multi-tiered strategy. Primary reaction wheel disturbances are minimized at the source through strict unbalance limits and further attenuated by a passive isolation system. The structural design actively avoids placing key modes near wheel harmonics, while operational protocols enforce accelerated pass-throughs to prevent amplification at critical resonances. A Fast-Steering Mirror (FSM) is commanded to suppress adverse LOS motion at or below wheel isolator frequency bands. Science data handling includes an X-band downlink to a network of commercial ground stations. The design supports multi-year mission operations with critical subsystem redundancy. 

% ----------------
\begin{figure*}[t!]
\centering
\includegraphics[width=0.99\textwidth]{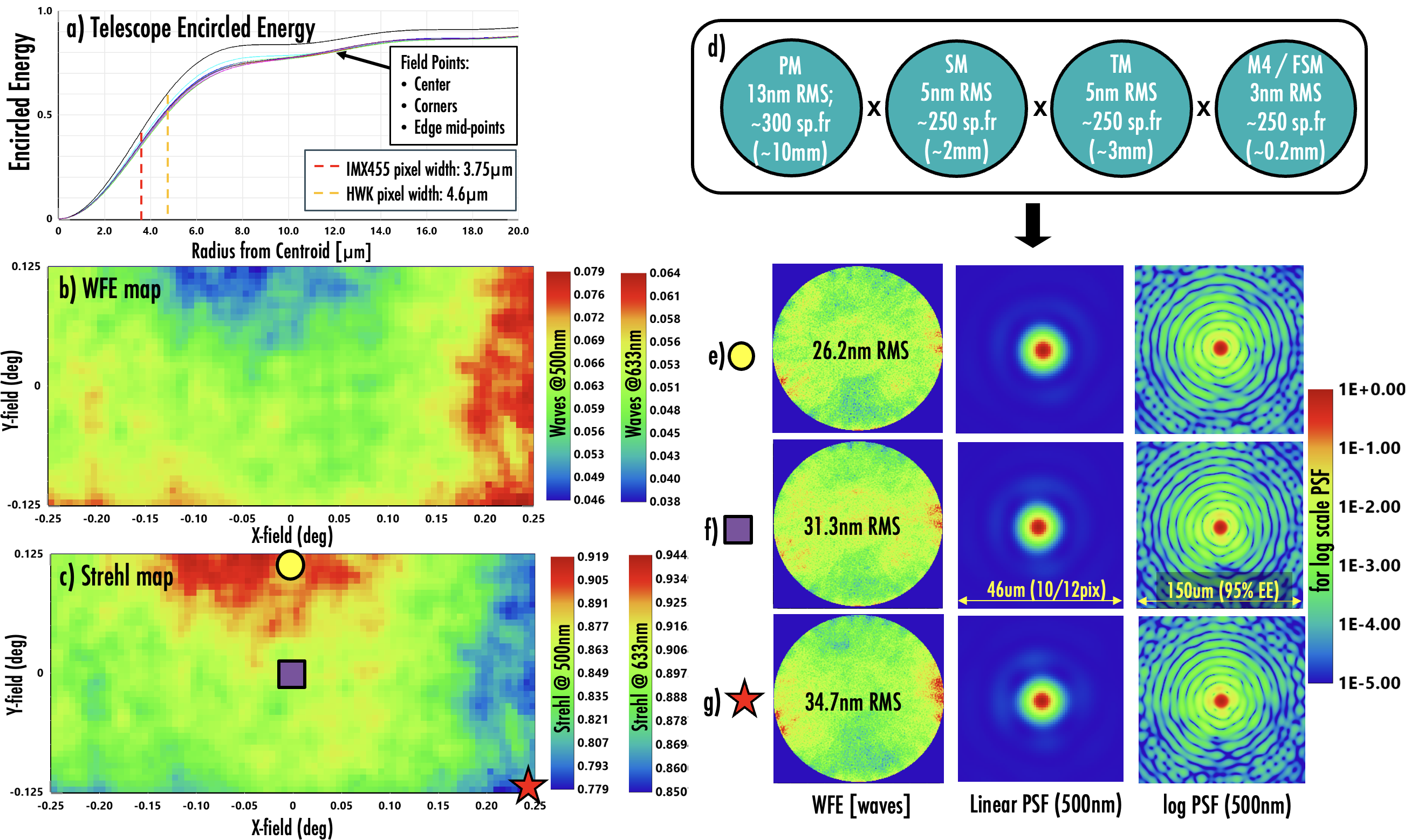}
\caption{Current model prediction for the as-built in-orbit image quality across the Lazuli focal plane. This accounts for surface figure errors across all four optical elements (PM, SM, TM, and FSM), residual alignment errors, and thermal distortion errors of the primary mirror (PM) of the TMA telescope. a) Encircled energy versus radius at various field points compared to the diffraction limit (black). Dashed vertical lines indicate the pixel size of the WCC sensors (red: Sony IMX 455; orange: HWK 4123). b) Wavefront error map in waves across the telescope focal plane with colorbars for 500 and 633 nm. c) Strehl ratio map of the same area. d) Surface figure assumptions for the four telescope optical elements---Primary Mirror (PM), Secondary Mirror (SM), Tertiary Mirror (TM), and FSM, showing the RMS wavefront and number of spatial frequencies (sp.fr.). The contribution of each of the four optics are multiplicatively combined to form the wavefront map, linear PSF, and log-colorbar PSF shown in e), f) and g). The circle, square, and star in c) indicate the field points which are shown in e), f) and g).}
\label{fig:wavefront}
\end{figure*}

\begin{figure*}[htb]
\centering
\includegraphics[width=0.98\textwidth]{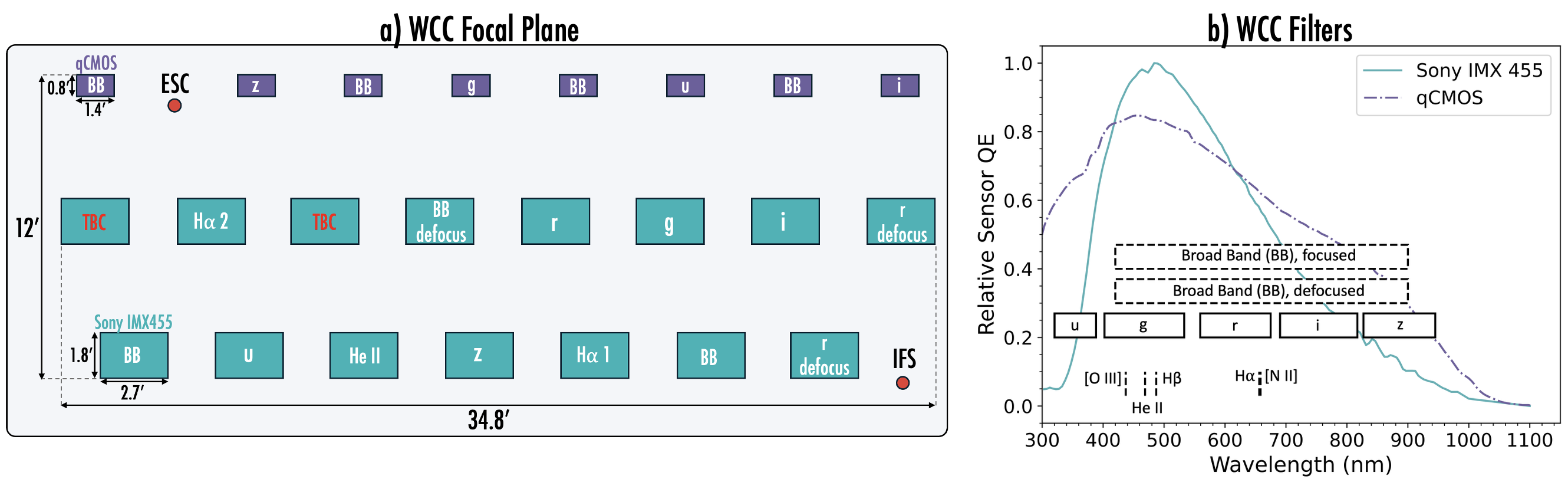}
\caption{a) Overview of the WCC focal plane showing the distribution of the sensors on the focal plane. Sony IMX sensors are shown in turquoise, and HWK qCMOS sensors are shown in purple. The entrance apertures to the IFS and ESC instruments are also indicated with the red circles. Filter locations are subject to change pending instrument design optimization. Possibilities for the `TBC' filter locations are discussed in the main text. b) Overview of the WCC filter suite as a function of wavelength. The nominal quantum efficiency of the Sony IMX 455 and the HWK 4123 qCMOS sensors are shown as solid black and dot-dashed green lines, respectively.}
\label{fig:wccfocalplane}
\end{figure*}

% ----------------------------------------------
\subsection{Optical Telescope Assembly}\label{sec:ota}
Lazuli is designed around an off-axis TMA telescope with an optically monolithic primary mirror (PM) of 3~m diameter. The PM construction follows a novel, proprietary approach, using lightweight, thermally stable, low coefficient of thermal expansion (CTE) materials including silicon carbide (SiC). The design traces heritage to commercially flying telescopes that deliver diffraction limited image quality in the visible, and also incorporates lessons learned from technologies developed for the JWST optics production. A key enabling factor behind Lazuli is its deep technological heritage drawn from these prior space programs and ground based telescopes, as well as using an active in-orbit alignment technology and jitter control. This informs the architectural choices and performance modeling underlying the telescope design, enabling Lazuli not only to meet its optical performance requirements but also to support the mission’s demanding development and deployment schedule.

The large, effectively unobscured, (kinematically) passive PM provides a unique opportunity for coronagraphy, while the freeform TMA design enables a large aberration-balanced field of view that provides diffraction limited PSFs to host multiple instruments. Controlling the telescope’s exit pupil on a fourth flat Fast Steering Mirror (FSM) enables jitter compensation which, along with the careful design of the bus and the attitude control system (ACS), is designed to meet tight pointing requirements enabling the observatory’s diffraction-limited image quality targets at 633~nm. The telescope structure largely relies on lightweight and low CTE structural materials. The selection of orbit, optical materials, concept of operations, and internal thermal design together ensure that Lazuli achieves both rapid and low-amplitude thermal settling between slews, once the payload and telescope structure have been aligned and stabilized on orbit during commissioning. Figure~\ref{fig:wavefront} shows expected model predictions for the as-built in-orbit image quality across the Lazuli focal plane.

% ----------------------------------------------
\subsection{Widefield Context Camera (WCC)}\label{sec:wcc}
The WCC is a diffraction-limited (Strehl $>$0.8 at a reference wavelength of 633~nm) wide field imager with a 35\arcmin$\times$12\arcmin\ footprint. The focal plane is populated with an array of 23 CMOS sensors (providing a sensor fill factor of $\sim$0.2, although there is an additional $\pm$15 degree flexibility in the telescope roll angle around the boresight that provides increased sky coverage), each equipped with a fixed photometric filter (Figure~\ref{fig:wccfocalplane}). Exact filter positions and arrangements are notional at this stage and subject to further optimization.

The WCC employs two detector types: the Sony IMX 455 CMOS sensor and the BAE HWK 4123 qCMOS sensor. The IMX 455 provides a larger field of view ($2.7\arcmin\times1.8\arcmin$ per sensor) and constitutes the majority of the array (15 sensors). The HWK 4123 offers a smaller field of view ($1.4\arcmin\times0.8\arcmin$ per sensor), but achieves sub-electron read-out noise (RON $<0.3e^-$), improving the limiting magnitude for low signal to noise (S/N) ($<10$) sources by $\sim$1.5~mag compared to the IMX 455 (RON $\sim 2e^-$). This low read noise is important to reach $>99\%$ guide star availability in the field of regard and fast closing of the FSM control loop. As such, the WCC is not just a science instrument but also provides core functionality to the observatory through guide-star observations. Each PSF is sufficiently sampled by multiple pixels and certain sensors are permanently offset in focus to provide service to other instruments (e.g., PSF knowledge for the IFS through phase retrieval wavefront sensing) or to enable the highest possible photometric precision by integrating over pixel-to-pixel and sub-pixel systematics.

The WCC will cover the visible wavelength range from the blue to the red sensitivity cut-offs of the silicon sensors, 350--1000~nm (Figure~\ref{fig:wccfocalplane} right panel). The filter set will include both broad- and narrow-band filters, including Sloan-like $u$, $g$, $r$, $i$, $z$ bands, along with a broad-band filter to enable high precision exoplanet transit science in integrated light. One of these filters will be for an in-focus sensor, while another for an out-of-focus sensor. Narrow-band filters will notionally include two H$\alpha$ filters of different widths (nominally 2nm and 6nm in width indicated as H$\alpha$-1 and H$\alpha$-2 in Figure \ref{fig:wccfocalplane}), along with a He\,\textsc{ii} filter. Additional possible narrow band filters are being considered for the two remaining slots indicated as 'To Be Confirmed (TBC)' in Figure \ref{fig:wccfocalplane}, including H$\beta$, [O {\sc iii}], and [N {\sc ii}].

In addition to full frame imaging, all detectors are capable of operating in region of interest windowed modes, nominally with frame rates up to at least 200 Hz. The combination of wide-field, low read noise detectors with a diffraction-limited 3~m aperture optical system  will serve a very broad range of science interests, including spatially resolved studies of extended sources and in crowded fields, high precision photometry, high temporal resolution ($\sim 5$ ms) science, and very faint source imaging.

The TMA design enables simple accommodation and low-risk operation of the WCC, with no moving parts required for the photometric sensors in a focal-plane layout intentionally configured to balance science, guiding, and wavefront-sensing requirements (see Figure~\ref{fig:wccfocalplane}). This design choice, however, requires multi-band observations of a given source to be obtained via telescope offsets that sequentially place the target on the desired sensors. Nevertheless, multiple sensors can be operated simultaneously to improve the efficiency of large mosaicking observations or to enable parallel investigations when target placement and guide-star availability permit. Such parallel observations can also be carried out when the WCC is not the primary instrument, enabling ancillary science programs—for example, the construction of {\it deep-field mosaics} as a by-product of repeated visits to the same region or deep integrations obtained concurrently with IFS observations.

\begin{figure}[h!]
\centering
\includegraphics[width=1\columnwidth]{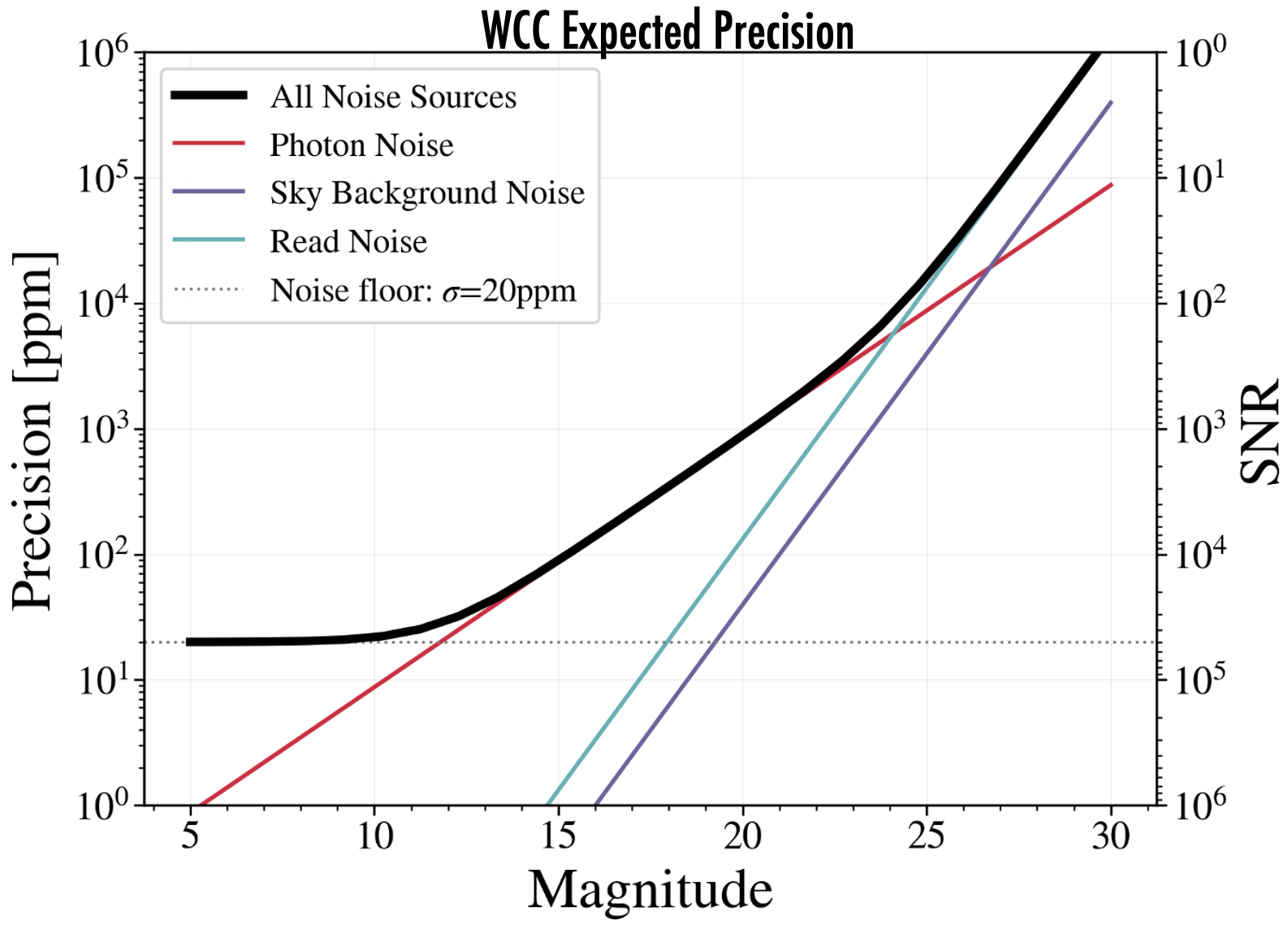}
\caption{Expected photometric precision (black curve; left axis) in ppm and S/N (right axis) as a function of stellar magnitude as observed in the in-focus WCC $r$ filter with the WCC in a 1~h effective exposure. The contributions from different noise sources are highlighted: photon noise (red), sky-background (purple), read noise (turquoise), and systematic noise floor (grey horizontal dashed line).}
\label{fig:wccprec}
\end{figure}

Lazuli's rapid response capability will make the WCC a transient workhorse, but dynamic scheduling is a more necessary capability for that science than pure photometric precision. The WCC noise floor will have the largest impact on transiting exoplanet observations, and we discuss the expected performance of the chosen architecture in this context. As an example of the expected on-sky performance of the WCC, Figure~\ref{fig:wccprec} shows the expected photometric precision (and S/N) of the WCC Sloan-like $r$ filter as a function of stellar magnitude for one of the in-focus $r$ band sensors. The precision estimates include contributions from photon, read, dark, and sky background noise. Observations of the brightest stars will be systematics limited---especially at long binning timescales---setting a systematic photometric precision noise floor due to contributions from a combination of imperfect detectors, calibrations, varying background light, and astrophysical sources of noise. In Figure~\ref{fig:wccprec}, we assume a noise floor of 20 ppm, comparable to the systematic noise floor achieved by Kepler for quiet solar-type stars on transit-relevant timescales \citep{gilliland2011}. The exact value of this noise floor needs to be refined, and ultimately tested on-sky. To enable precise photometric observations of exoplanet transits, one of the sensors will have a broad-band Kepler-like bandpass filter (nominally from 400--900~nm) to maximize the stellar flux rate, while being defocused slightly to enable better averaging over inter-pixel sensitivity effects. For the high photometric precision observations, care will be taken to enable the capability of observing and extracting data of nearby reference stars, leveraging lessons learned from the Kepler and TESS missions on co-trending basis vectors \citep[e.g.,][]{stumpe2012keplerPDC,jenkins2016STESSSOC}.Exposure time calculator and tools have been developed for the WCC and will be made available to the community in the near future.

\begin{figure*}
\centering
\includegraphics[width=0.98\textwidth]{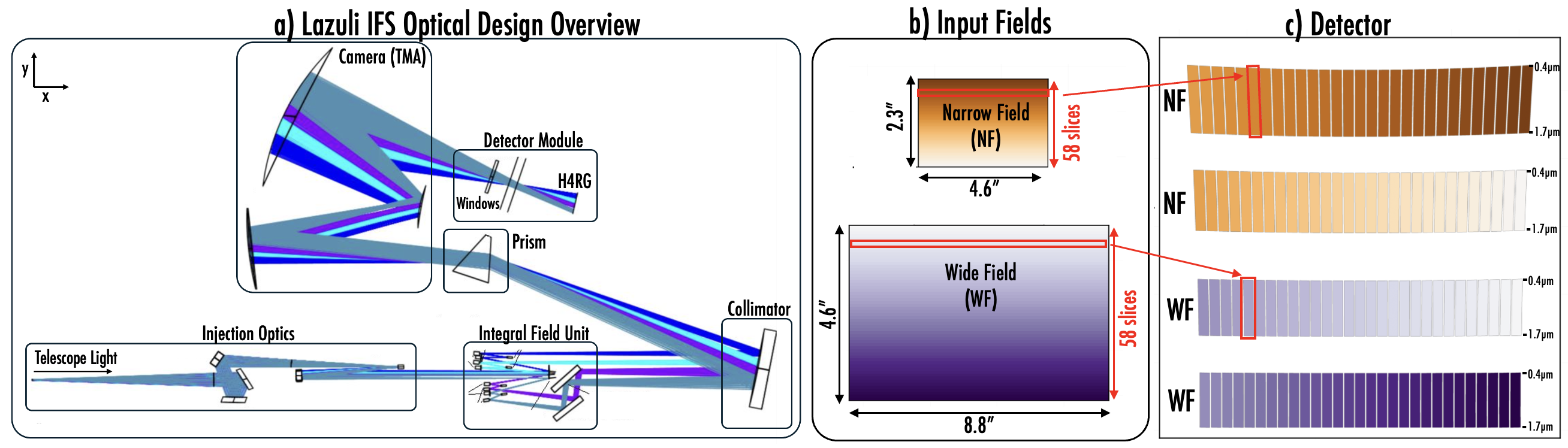}
\caption{a) Overview of the optical design of the Lazuli Integral Field Spectrograph, showing the injection optics, integral field unit (IFU), collimator, prism, the camera subsystem, and the detector module subsystem. b) The input fields of view of the IFS, which is composed of a 2.3$\times$4.6\arcsec~Narrow Field (NF), and a 4.6$\times$8.8\arcsec~Wide Field (WF). c) Overview of the expected locations of the 116 traces (58 per subfield) as viewed on the $4096 \times 4096$ pixels of the H4RG-10 detector. The overview of the slices is generated by the \texttt{slicersim} code (Rigault et al. 2026, \textit{in prep.}).}
\label{fig:ifs}
\end{figure*}

% ----------------------------------------------
\subsection{Integral Field Spectrograph (IFS)} \label{sec:ifs}
The IFS provides continuous wavelength coverage from 400--1700~nm---spanning more than two octaves---at spectral resolution $R \sim 100-500$. The IFS offers two parallel fields: a Narrow Field (NF) with a 2.3\arcsec$\times$4.6\arcsec\ field of view with  40~milli-arcsecond (mas) spatial sampling, and a Wide Field (WF) with a 4.6\arcsec$\times$8.8\arcsec\ field of view with 80~mas spatial sampling. 
The IFS has minimal moving parts for science observations to increase its reliability and calibration consistency.

The optical design builds on slicer integral field designs from SNAP \citep{aldering2002, ealet2006}, WFIRST/Roman \citep{gao2017}, and, most directly, ORKID-II \citep{orkidii2024}. It consists of foreoptics, an image slicer integral field unit (IFU), and a TMA spectrograph. The foreoptics relay the telescope focal plane to a reimaged focus at the IFU entrance, creating a slow beam to accommodate the slicer geometry and introducing 1:2 anamorphism to increase the signal-to-noise ratio on the detector for faint continuum-source objects. A pupil image within the foreoptics enables calibration injection matched to the telescope illumination. The IFU employs a diamond-turned aluminum image slicer, with heritage from the DKIST solar telescope \citep{Anan2024MISI36} and the INFUSE rocket-borne mission \citep{witt2021}. The slicer consists of 58 slices for each of the NF and WF sky fields, reformatting the two-dimensional field into a pseudo-slit. The spectrograph uses an off-axis parabola collimator, a prism disperser for continuous wavelength coverage without order overlap, and a TMA to reimage the dispersed spectra onto the detector.

The detector is a Teledyne H4RG-10 with a 1700~nm cutoff, selected for its performance across the broad IFS bandpass. The median quantum efficiency exceeds 50\% at 800~nm and 70\% at 1200~nm, with goals of 60\% and 80\% respectively. The detector is passively cooled to an operating temperature of 120~K to minimize dark current, and is expected to achieve median dark current $\leq$0.01~e$^-$/s/pix with a goal of $\leq$0.001~e$^-$/s/pix. The median correlated double sampling (CDS) read noise is $\leq$25~e$^-$, with a goal of 20~e$^-$.

The IFS is designed for high-precision spectrophotometry spanning up to four orders of magnitude in flux, enabled by comprehensive 2D and 3D calibration systems. The 3D calibration system injects light at the foreoptics pupil plane to match the telescope illumination, providing flat fields through the full spectrograph optical path. Wavelength calibration is achieved using a quartz-tungsten-halogen (QTH) lamp combined with a Fabry-Perot etalon and laser diode reference line. The 2D calibration system illuminates the detector directly via four LED sources spanning the IFS passband, enabling flux-dependent linearity corrections across the full dynamic range and characterization of detector regions receiving low flux from the spectrograph optics, such as near the gaps between slice projections. A calibration shutter mechanism enables dark exposures without thermal contributions from the telescope or sky. Together, these systems substantially reduce or eliminate the need for on-sky flat-field observations, which require astrophysical sources that are rarely sufficiently uniform for the precision required.

% ----------------------------------------------
\subsection{ExtraSolar Coronagraph (ESC)}\label{sec:esc}
\begin{figure*}[t!]
\centering
\includegraphics[width=0.99\textwidth]{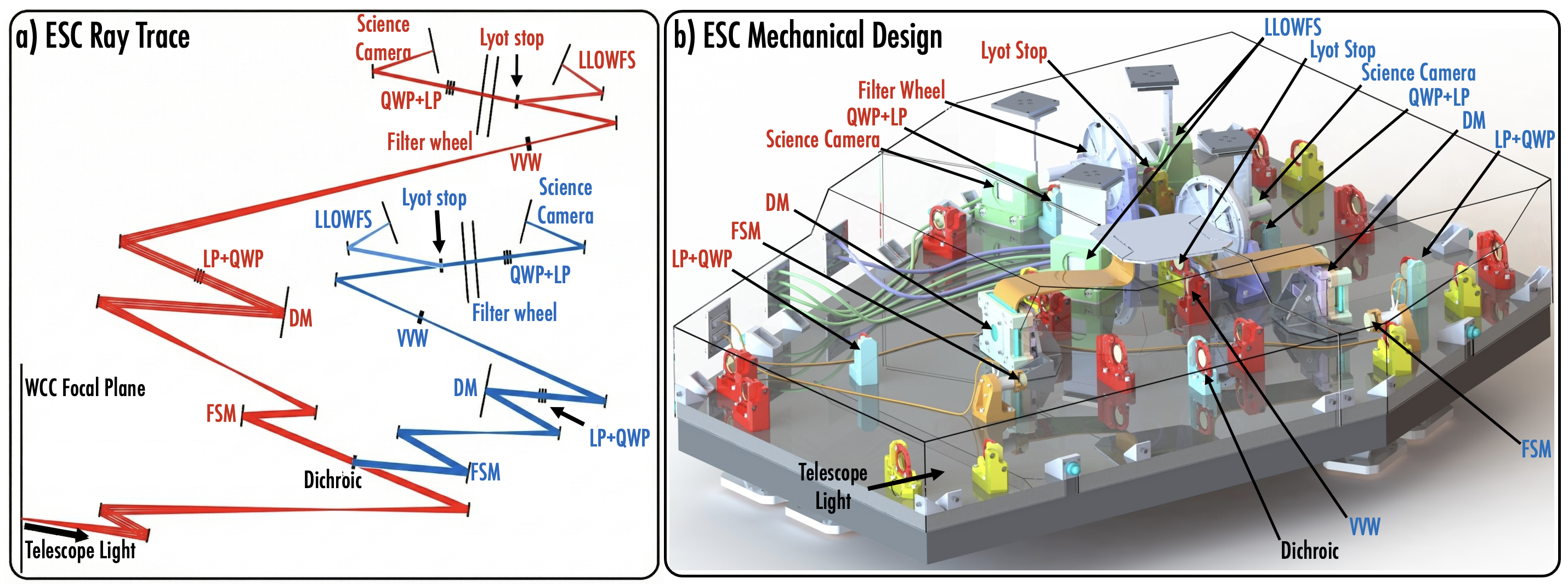}
\caption{Preliminary optical ray trace and mechanical design of the ESC instrument. a) Light enters from the bottom left where a dichroic  splits the light into a red and a blue channel, followed by a piezo-electric Fast Steering Mirror \citep[PZT FSM;][]{mendillo_flight_2012} for fine guiding, 1K and 2K Boston Micromachines MEMS deformable mirrors for active wavefront control, Linear Polarizer (LP) and Quarter Wave Plate (QWP) for polarization filtering, charge-6 liquid-crystal polymer Vector Vortex Waveplate (VVW) coronagraph masks, reflective Lyot-stops to feed a Lyot Low-Order Wavefront Sensor \citep[LLOWFS;][]{mendillo_reflective_2023},  optional self-coherent camera modes \citep{derby_integrated_2023}, selectable narrow-band filters, and low-noise CMOS science cameras. b) A composite optical bench supports ruggedized optics mounts, adapted from \cite{huie_design_2024}, and a filter wheel is adapted from OSIRIS-REx OCAMS \citep{rizk_ocams_2018}. Baffles, covers, electronics boxes are not shown and components are artificially colored for emphasis. Computer-aided design (CAD) figure credit: H. Olivas and G. West. Coronagraph design concept developed by the University of Arizona.}
\label{fig:escpanel}
\end{figure*}

The ExtraSolar Coronagraph (ESC; see Figure~\ref{fig:escpanel}) is a high-contrast imaging system designed to enable direct imaging and characterization of exoplanets and circumstellar debris disks around nearby stars. The ESC design relies on many recent technological advances, leveraging progress in adaptive optics (AO) and  coronagraph technology development on ground-based observatories (see review in \citealt{deeg_direct_2018}), suborbital coronagraph missions \citep{mendillo_flight_2012,douglas_wavefront_2018,mendillo_first_2020}, and laboratory testbeds (for a summary review, see \citealt{mennesson_current_2024}). Particularly influential in the Lazuli ESC design have been the PICTURE-C mission \citep{cook_planetary_2015,mendillo_balloon_2023}, the DeMi CubeSat \citep{douglas_small_2021,morgan_optical_2021}, work at NASA's High Contrast Imaging Testbed Facility at JPL \citep{trauger_laboratory_2007,ruane_broadband_2022,potier_random_2023}, and the University of Arizona Space Coronagraph Optical Bench \citep{maier_design_2020,kim_advances_2021,ashcraft_space_2022,van_gorkom_space_2022,ashcraft_space_2024,van_gorkom_space_2024}. The science yield of the high-contrast design will then be mapped leveraging community developed tools such as  EXOSIMS, the Exoplanet Open-Source Imaging Mission Simulator  \citep{savransky_exosims:_2017}. These efforts provide a strong technical foundation and substantially reduce development risk.

Building on these foundations, the Lazuli ESC employs a two-arm coronagraph design spanning 400--750 nm that requires minimal new technology development. The system combines 1K and 2K micro-electromechanical system (MEMS) deformable mirrors \citep{bifano_continuous-membrane_1997,douglas_wavefront_2018,potier_random_2023}, charge-6 vector vortex wave plates in the focal plane \citep{serabyn_technology_2019}. Together with a Lyot stop the vortex wave plate forms a vector vortex coronagraph (VVC) \citep{mawet_vector_2010, ruane_performance_2017,ruane_broadband_2022}.  Finally, CMOS detectors, and a software architecture derived from the MagAO-X instrument \citep{males2024magAOX} which shares many common packages with other advanced ground-based AO projects \citep{guyon_compute_2018,skaf_real-time_2024} provide sensing and control to suppress speckles. Running proven AO software in Linux on industrial embedded computers with onboard graphic processing unit (GPU) acceleration greatly enhances the flexibility and shortens the time to deployment versus a traditional flight software development life cycle. 

\begin{figure*}[t!]
\centering
\includegraphics[width=1\textwidth]{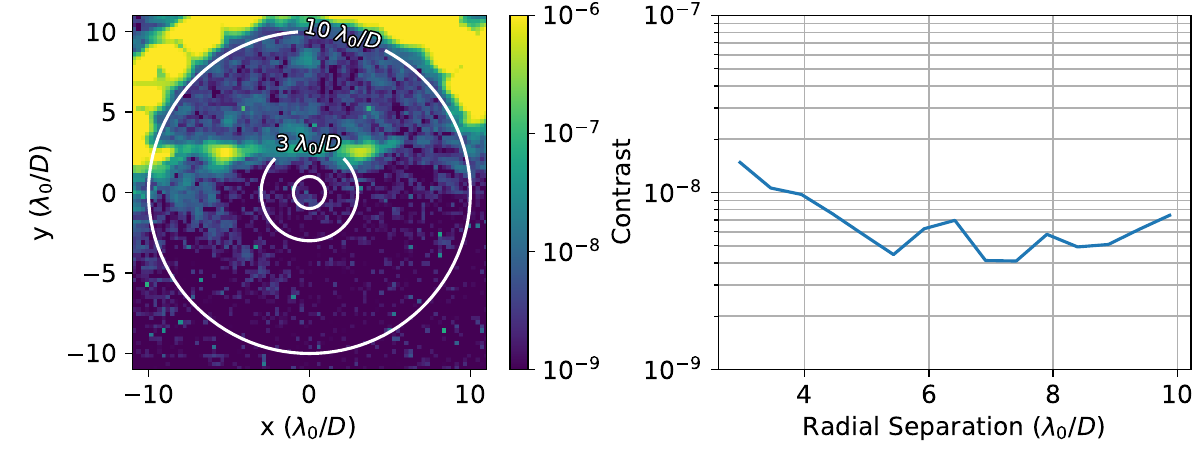}
\caption{Measured contrast on the Space Coronagraph Optical Bench (SCoOB) testbed in vacuum in a 2\% bandwidth centered at $\lambda_0=630$nm. The mean normalized intensity contrast in a D-shaped dark hole from 3-10$\lambda_0/D$ is $5.8\times10^{-8}$. Adapted from \cite{van_gorkom_space_2024}.}
\label{fig:scoob_vac_2pcent_630nm}
\end{figure*}

Wavefront sensing and control are implemented using a two-fold approach to address a wide range of spatial and temporal frequencies. First, low-order aberrations, including pointing errors, are corrected using a Lyot Low-Order WaveFront Sensor (LLOWFS) that uses light diffracted by the vortex focal plane masks and reflected from the Lyot stop \citep{singh_-sky_2015}, an approach that has been optimized and demonstrated in relevant regimes \citep{mendillo_reflective_2023,milani_space_2025}.
Even a well corrected optical system with a VVC will have speckles of order 10$^{-5}$ due to manufacturing errors; thus a second, focal plane speckle suppression step is required. This is achieved using half-focal plane High Order WaveFront Sensing (HOWFS) to generate a `` dark hole'' region of high contrast where the speckle intensity due to phase and amplitude errors are minimized \citep{giveon_broadband_2007}; see for example Figure~\ref{fig:scoob_vac_2pcent_630nm}. We have baselined the proven implicit-Electric-Field conjugation technique \citep{haffert_implicit_2023,milani_simulating_2023} which minimizes sensitivity to model errors and has been shown to provide four orders of magnitude of speckle suppression in 2\% bandwidth with a comparable coronagraph \citep{van_gorkom_space_2022,van_gorkom_space_2024}. Other optional HOWFS modes to measure and/or stabilize the PSF are included, such as linear-dark-field control \citep{miller_spatial_2017} and a self-coherent camera mode for spatial variation of the speckle field in the dark hole \citep{baudoz_self-coherent_2005,potier_comparing_2020,derby_space_2025}. The SpaceVPX onboard compute system based on NVIDIA AGX Orin allows HOWFS operations to be performed onboard. The half-dark-hole approach provides high-contrast and throughput without requiring the stringent high spatial frequency static surface smoothness of two-DM-in-series designs \citep{mazoyer_fundamental_2017}, at the expense of a concept of operations which requires multiple separate observations to perform $360^\circ$ imaging around a single star. The onboard Linux computing environment further enables testing of alternative HOWFS algorithms, including adjoint methods \citep{milani_demonstrations_2025}, linear dark-field control \citep{miller_linear_2016}, and related approaches.

\begin{figure*}[t!]
\centering
\includegraphics[width=1\textwidth]{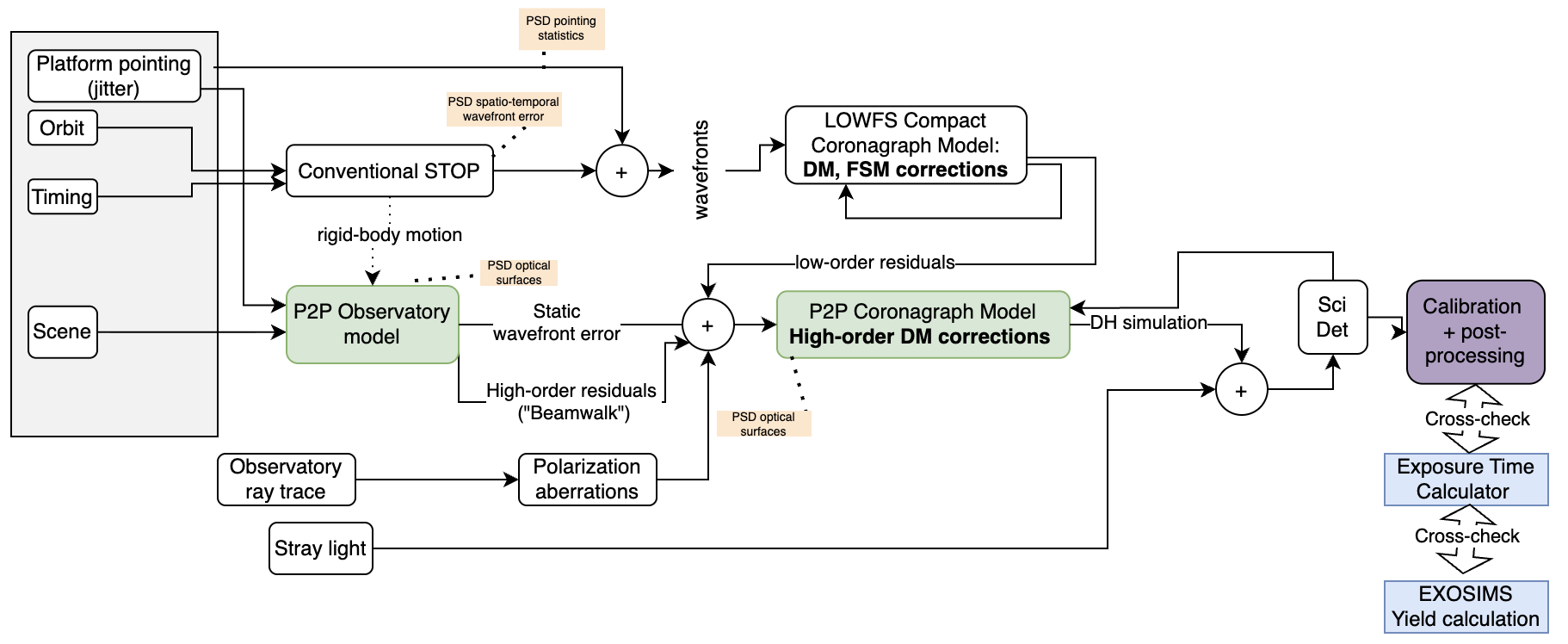}
\caption{Planned simulation flow for modeling of  Lazuli coronagraphic observations. Inputs left, define the pointing environment, orbit, timing, and scene which define behavior of a Structural Thermal Optical Performance (STOP) analysis and the inputs to and plane to plane (P2P) paraxial diffraction model. A LOWFS model defines what platform dynamic residuals can be controlled which are then provided as inputs to the high-order DM corrections model which includes HOWFS, the MEMS DM behavior and wavefront surface errors, outputs are combined as an intensity incident on the Science Detector (Sci Det) which then can be used as an input to post-processing simulations which can then be checked against exposure time estimates or used as inputs to the next iteration of the HOWFS model.}
\label{fig:sim_flow}
\end{figure*}

Early testing of this approach has demonstrated contrasts better than $10^{-8}$ in 5\% or narrower bandwidths using existing hardware \citep{van_gorkom_space_2024,milani_demonstrations_2025}. An in-vacuum laboratory-measured dark hole from the SCoOB testbed, in a configuration analogous to the red channel of Lazuli, is shown in Figure~\ref{fig:scoob_vac_2pcent_630nm}. Combining the coronagraph with an unobscured 3-meter-class telescope opens significant discovery potential for both exoplanet imaging and circumstellar disk studies while demonstrating several new technologies. To support performance prediction and system optimization, the team has been developing a range of tools to accelerate the end-to-end modeling framework, including GPU-accelerated angular spectrum methods for high-contrast imaging modeling \citep{milani_modeling_2024} and post processing \citep{krishnanth_deepest_2024} using CuPy \citep{okuta_cupy_2017}, extending \texttt{Batoid} \citep{meyers_batoid_2019} for parallelized C++ ray tracing of complex surfaces for optical tolerancing to enable STOP modeling (Nicolas et al.~2026, \textit{in prep}), and PyTorch implementations of Karhunen–Lo\'{e}ve post-processing \citep{ko_pytorch_2024}. A single end-to-end contrast budget is used to track the contribution of instrument and observatory systematics building on prior work from many teams \citep{ndiaye_high-contrast_2013,mendillo_optical_2017,nemati_sensitivity_2017,nemati_analytical_2023,van_gorkom_performance_2025-1}. End-to-end simulations provide a means of verification and validation of the contrast budget terms and the planned coronagraphic observatory simulation flow (Figure~\ref{fig:sim_flow}) inspired by the Roman Coronagraph modeling approach \citep{krist_wfirst_2018}, using power spectral density (PSD) representations of spatiotemporal error distributions as described in \citet{douglas_approaches_2023}. Initial results of this approach will be published in Douglas et al.~2026 \textit{(in prep)}.  As the observatory design coalesces, statistically defined surfaces and temporal variations will subsequently be updated and/or replaced with as-built measured optical surface error maps and STOP time series results.

% ----------------------------------------------
\section{Science Capabilities \& Touchstone Use Cases}\label{sec:science}
The Lazuli Space Observatory is designed with an instrument suite that enables access to regions of observational parameter space that remain poorly explored, particularly at the intersection of rapid response, stable spectrophotometry, and broad optical--near-infrared wavelength coverage. In this section, a set of \emph{touchstone use cases} is presented to illustrate how Lazuli's design capabilities translate into high-impact science. The examples are organized around three high priority science areas for Lazuli: time-domain and multi-messenger astronomy (\S~\ref{sec:tdamm}), stars and planets (\S~\ref{sec:exoplanets}), and cosmology (\S~\ref{sec:cosmology}); they were selected both for their scientific importance and for the role they played in shaping key observatory requirements. While not exhaustive, these use cases highlight the breadth of studies enabled by Lazuli, ranging from high-cadence ($\sim$5~ms) photometry with the WCC (e.g., X-ray binary time-lag measurements, searches for fast optical counterparts to fast radio bursts, and pulsar studies) to complementary cosmological probes (e.g., an independent $H_0$ measurement from strongly lensed supernovae with the IFS) and multi-faceted investigations of planet formation and evolution pathways.

% ----------------------------------------------
\subsection{Time-Domain and Multi-Messenger Astronomy}
\label{sec:tdamm}
When Lazuli starts science operations, it will join a suite of facilities dedicated to the pursuit of time-domain and multi-messenger astronomy. 
 
This will include several wide-field surveys designed to systematically explore, for the first time, very short timescales (minutes -- a day), including the Argus array (optical, \citealt{Law2022}), Rubin's LSST combined with other surveys such as the La Silla Schmidt Southern Survey (optical, \citealt{2025PASP..137i4204M}), ULTRASAT (UV, \citealt{2024ApJ...964...74S}), and the Deep Synoptic Array (DSA; radio, \citealt{2019BAAS...51g.255H}). 

To fully exploit the scientific opportunities that these facilities will generate, including detailed studies of the earliest phases of transient evolution as well as the characterization of rare and/or currently unknown phenomena, a key bottleneck in the existing infrastructure is the capability to rapidly obtain follow-up photometry and spectroscopy. 

To probe the evolution of fast-evolving transients in detail, a combination of rapid-response and sensitivity afforded by extremely large ground-based, or large space-based observatories, is required. While planned/proposed missions will cover this combination of capabilities at other wavelengths (e.g., UVEX, \citealt{2021arXiv211115608K}; AXIS, \citealt{2023SPIE12678E..1ER}), no $\gtrsim 1$~m space-based observatory planned in the next decade covers optical/NIR spectroscopy. 

Lazuli, with its large aperture and a mission operations concept designed to be flexible and responsive, is expected to enable follow-up observations of the faintest, fast-evolving transients at scale, thereby opening up a new part of parameter space for systematic exploration. Lazuli’s concept of operations is structured to enable response times to community Target of Opportunity (ToO) requests on timescales shorter than 4 hours (from trigger submission to open shutter on target), with a goal of 90 minutes. In addition to rapid-response, Lazuli will have the capability to perform high-cadence (down to $\sim$ hours, in principle) monitoring over timescales of weeks with minimal interruptions, providing regular sampling of transient evolution that is very hard to reliably achieve from the ground.

% ----------------------------------------------
\subsubsection{Fast transients}
\begin{figure}
\centering
\includegraphics[width=0.9\linewidth]{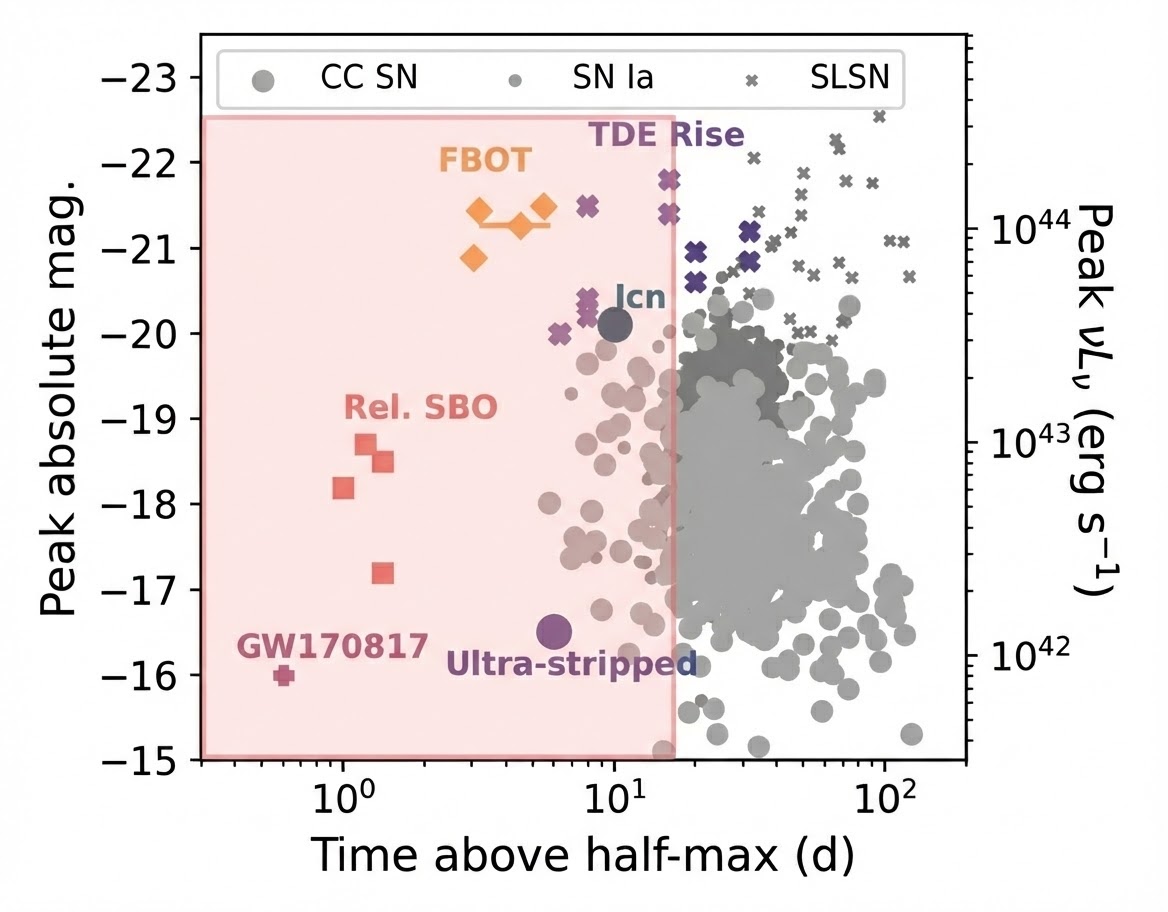}
\caption{Lazuli provides new opportunities to study faint and fast-evolving phenomena. The red box highlights where Lazuli's rapid response and sensitivity will open up new parameter space for systematic exploration. Even for the faint end (absolute magnitude of --15), Lazuli's distance horizon for spectroscopy of fast-evolving transients is $\lesssim1$ Gpc. Highlighted sources include GW170817, the kilonova counterpart to a binary neutron star merger; relativistic supernova shock-breakout (Rel.~SBO) events; fast blue optical transients (FBOT); and rapidly evolving (ultra-)stripped envelope supernovae such as SNe\,Icn. {\it Normal} supernovae are shown in grey. The \textit{x-axis} shows how much time a source spends above half of its maximum brightness, a proxy for whether its evolution is fast or slow. Figure adapted from \citet{2021arXiv211115608K}. }
\label{fig:rapidresponse}
\end{figure}

Over the past two decades, a range of fast-evolving astrophysical phenomena with characteristic timescales of milli-seconds to weeks has been discovered. These include fast radio bursts; optical and infrared flaring from X-ray binaries and magnetars; luminous, UV and X-ray-bright fast transients that potentially could be powered by the tidal disruption (and in some cases detonation) of a white dwarf around an intermediate-mass black hole (see Gezari et al.~2026, \textit{in prep.}, for a detailed analysis); and fast transients such as supernova shock breakout (SBO) and jet-driven events from massive stars, among others (Figure~\ref{fig:rapidresponse}). Despite growing interest, the number of well-characterized sources remains small, and in several cases robust classifications have yet to be established.

These events occupy an extreme corner of parameter space: they are intrinsically rare, distant, and often too faint and/or short-lived for systematic multi-wavelength follow-up. With its combination of rapid response, deep optical/NIR imaging and low-resolution spectroscopy, and flexible scheduling, Lazuli will overcome the main bottlenecks limiting exploration of this parameter space, enabling both routine classification and detailed physical interpretation of these rare phenomena. For reference, in a 6 hour observation with the IFS Lazuli can deliver S/N$>$5 over most of the covered wavelength range for a peak absolute magnitude of --15 at a distance of $\sim$1 Gpc ($z\sim 0.2$). For a peak absolute magnitude of --22, the distance horizon for spectroscopy extends out to $z \sim 3$.

In addition to the exploration of this poorly understood part of parameter space, we highlight a number of science cases where Lazuli is expected to provide new insights and/or highly complementary capabilities compared to existing facilities. These scenarios showcase the power and potential of Lazuli to improve our understanding of the dynamic Universe and have a broad impact on time-domain science.

% ----------------------------------------------
\subsubsection{Gravitational Wave Follow-Up}
Gravitational wave detections from merging neutron stars (NSs) and black holes (BHs) have opened up a new window on the Universe. The most likely detectable electromagnetic counterpart to a binary NS merger (BNS) is a fast-evolving, faint transient with potential emission from $\gamma$-rays through X-ray, UV, optical, NIR, sub-mm and radio wavelengths. This emission encodes fundamental information on the physics of dense matter, the formation of heavy elements, and the nature of the merger remnant. In addition, NS-BH mergers may also produce fast-evolving and faint kilonova emission \citep{2025PhRvD.112l3005K}; this emission has not yet been detected but occupies the parameter space in which Lazuli will excel.

\begin{figure}
    \centering
    \includegraphics[width=\linewidth]{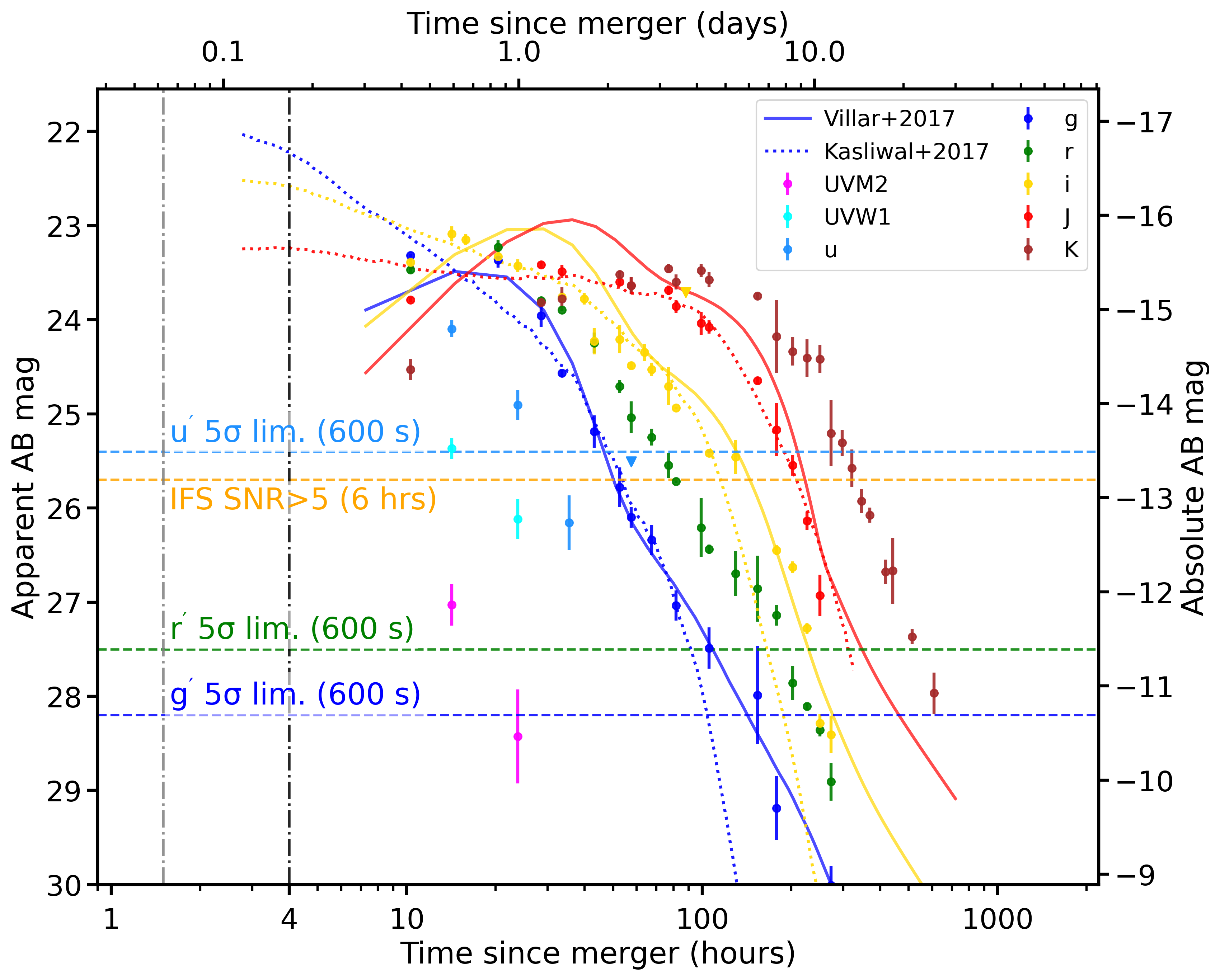}
    \caption{Multi-band lightcurves of the kilonova counterpart to the binary neutron star merger GW170817, shifted to a distance of 600 Mpc. Lazuli's rapid response will enable very early spectroscopic and photometric constraints of future kilonovae, where the model predictions diverge and hence discriminating power is largest. It can obtain broad band lightcurves and spectroscopy for nearly all BNS mergers out to 1--1.5 Gpc, provided that the correct counterpart is identified in a timely manner. The vertical dash-dotted lines show the 4 hour requirement (black) and 90 minute goal (grey) for ToO response time. Horizontal colored lines indicate the 5$\sigma$ limiting magnitude for WCC photometry in the $u$, $g$ and $r$ bands (light blue, dark blue, and green respectively), and the limiting magnitude to obtain a S/N$\gtrsim$5 spectrum with the IFS (orange).}
    \label{fig:kilonova}
\end{figure}

We show the multi-band lightcurves of the kilonova counterpart to GW170817 from \citet{2017ApJ...848L..17C, 2017ApJ...848L..19C, 2017ApJ...848L..24V, 2017ApJ...848L..27T, 2017Natur.551...64A, 2017Natur.551...67P, 2017Natur.551...71T, 2017Natur.551...75S, 2017PASA...34...69A, 2017PASJ...69..101U, 2017Sci...358.1559K, 2017Sci...358.1565E, 2017Sci...358.1570D,soares-santos2017EMcounterpartGW170817} in Figure~\ref{fig:kilonova}, shifted to a distance of 600 Mpc (which is the median expected distance for BNS mergers for which counterparts will be detectable in the fifth LIGO/Virgo/KAGRA observing run, e.g., \citealt{2025ApJ...993...15K}), together with estimates for the limiting magnitudes of Lazuli IFS spectroscopy and WCC imaging. We also overplot two example kilonova models (the radioactive decay model of \citealt{2017Sci...358.1559K} and the shock-cooling + boosted radioactive decay model of \citealt{2017ApJ...851L..21V}) to highlight the discriminating power of early observations. 

Lazuli can provide continuous 400--1700~nm spectroscopy for classification and characterization from the first hours up to $\sim$7 days post-merger at 600 Mpc. The distance horizon for Lazuli spectroscopy for a GW170817-like kilonova at peak is $\approx$ 1 (1.5) Gpc; this volume encompasses $>$85\% of all BNS mergers expected to be detected by the Ligo-Virgo-Kagra observatory during its next O5 observing run \citep{2025ApJ...993...15K}.

The most model-constraining phases of kilonovae occur within the first hours to day after the event \citep{2018ApJ...855L..23A}, when the emission is rapidly evolving, relatively blue, and directly shaped by the composition, velocity, and geometry of the ejecta. Capturing this early light requires the combination of fast response, regular high-cadence monitoring and sensitivity that no existing or planned space observatory is designed to provide at scale.

Lazuli's rapid response will provide constraining power to differentiate between different kilonova components such as shock-heated material, disk winds, and lanthanide-rich ejecta, thereby constraining the production of the heaviest elements in the Universe \citep[e.g.,][]{2017Natur.551...80K, 2020LRR....23....1M}. Lazuli’s broad wavelength coverage and stable spectrophotometry are ideally suited to identifying key spectral features, including potential signatures of r-process species at wavelengths that are inaccessible from the ground. At later times (days to weeks), Lazuli can track the potential emergence of the afterglow from an off-axis jet, providing crucial information about jet structure, viewing angle, and the physics of relativistic outflows (e.g., \citealt{2017ApJ...834...28N}). Together, and in combination with other multi-wavelength rapid-response facilities, these measurements will enable a detailed reconstruction of the merger: from the composition of the ejecta and the fate of the remnant to the geometry and energetics of any associated jet. A more detailed analysis of Lazuli's capabilities for multi-messenger astronomy will be presented in Kunnumkai et al.~2025 (\textit{in prep.}).

% ----------------------------------------------
\subsubsection{Fast Blue Optical Transients}
Luminous fast blue optical transients (LFBOTs, e.g., \citealt{Prentice2018}) further exemplify the scientific opportunities with Lazuli. These events are characterized by bright ($>$10$^{43}$ erg s$^{-1}$) emission, spanning radio through X-ray bands (e.g., \citealt{2019ApJ...872...18M}). Their rapid evolution, including continuum cooling, the emergence (or lack thereof) of spectral features indicating high velocity ejecta and/or shock interaction features (e.g., \citealt{2019ApJ...872...18M, 2019MNRAS.484.1031P}), hold clues to their progenitors, the explosion geometry, and the nature of the (tentative) central engine.

The 400--1700~nm IFS coverage will enable detailed tracking of the optical/NIR spectral energy distribution (SED) evolution, while simultaneously capturing the emergence and temporal evolution of spectral features. Together with UV from facilities such as HST, X-ray and radio observations, Lazuli can provide the high cadence, panchromatic data required to break model degeneracies. One illustrative example is the persistent near-infrared excess observed in AT2018cow, whose origin remains uncertain: it has been attributed to either i) dust echoes of circumstellar material -- offering insight into the mass-loss history and nature of the progenitor \citep[e.g.,][]{2023ApJ...944...74M} -- or ii) to reprocessing by a dense outflow, which could constrain outflow geometry, energetics, and the nature of the central engine (e.g., \citealt{2025ApJ...991..180C}).

Equally transformative is Lazuli’s ability to characterize the minute-scale optical flares such as those recently discovered in the LFBOT AT2022tsd \citep{2023Natur.623..927H}, with a typical peak magnitude of --20 (AB mag) in the optical bands. Lazuli will have the sensitivity to measure flare duty cycle, energetics and substructure on 10s of seconds timescales with the WCC. In addition, it will be capable of time-resolved spectroscopy on 1-2 minutes timescales to detect color changes and continuum shape variations for sources out to $z \sim 1$ (S/N $\gtrsim$ 5 across the wavelength range), providing direct constraints on the origin of this emission (synchrotron, magnetar-powered, or jet-driven). A more detailed summary of the wide range of time-critical and transient science cases that Lazuli's capabilities will enable can be found in \citet{2026arXiv260617136W}.

% ----------------------------------------------
\subsection{Stars and Planets}\label{sec:exoplanets}
Planets and their host stars evolve in tandem, from the earliest stages of planet formation through main-sequence evolution, potential habitability, and eventual dynamical or radiative disruption. The following subsections describe the science considerations that most strongly influenced the selection and design of Lazuli’s instrument capabilities for characterizing stars, exoplanets, and our own Solar System. These investigations are expected to make use of all three Lazuli instruments (ESC, WCC, and IFS), operating in complementarity with \emph{TESS}, \emph{JWST}, \emph{HST}, Roman, PLATO, Ariel, and other current and upcoming facilities.

% ----------------------------------------------
\subsubsection{Direct Imaging of Habitable Zones, Giant Planets, and Circumstellar Disks with the ESC}
The drive to deepen our understanding of Earth’s history, climate, and uniqueness in the Universe motivates the search for other stellar systems and a broader understanding of the Solar System’s context within the Galaxy. Exploring the formation, composition, and dynamics of planets leads to general conclusions about the occurrence rates of exoplanets and robust physical measurements of specific planets that test local models. Large samples are required for statistically robust measurements, such as the occurrence rate of short period planets around FGK stars inferred by Kepler transit observations (e.g., \citealt{winn_occurrence_2015,kunimoto_searching_2020}, and many others) or wide-orbit planet occurrence rates measured by microlensing surveys such as the Optical Gravitational Lensing Experiment  \citep[OGLE;][]{poleski_wide-orbit_2021} or the upcoming Roman Microlensing Survey \citep{penny_predictions_2019,boss_roman_2025}. 

However, getting to large numbers of planets requires searching around dim, distant stars in addition to bright, nearby hosts. Such surveys do not tend to discover suitable planets for follow-up observations that probe physical properties at the spatial scales of a planetary radius or temporal scales shorter than a human lifespan. Transit surveys of brighter stars (e.g., with TESS; \citealt{guerrero_tess_2021}) provide a better sample for follow up with transit spectroscopy, while radial velocity and direct imaging surveys of nearby stars are the most direct ways to find planets that can be characterized in detail. Ground based observations with 5-10 m class telescopes equipped with AO have resolved young, warm, freshly formed or adolescent giant planets in emission \citep[e.g.,][]{bowler_imaging_2016} as well as large numbers of bright circumstellar disks \citep[e.g.,][]{avenhaus2018, esposito_debris_2020}. Extending extreme adaptive optics technology to the upcoming 30m class telescopes \citep{guyon_extreme_2018, fitzgerald_planetary_2022, jensen-clem_updated_2022,chauvin_direct_2023, males2024GMagAO-Xpreliminary} is expected to lead to the imaging of Earth-like planets around nearby M dwarf stars. For most hypothetical exoplanets around FGK stars, reflected light is $10^7-10^{10}$ times dimmer than the host star (the ``star-planet flux ratio'') and circumstellar debris disks span an even larger dynamic range of resolution element to host star contrast. The Nancy Grace Roman Space Telescope Coronagraph, expected to launch in late 2026, with multiple active optics, is likely to take our first image and spectrum of a Jupiter-analog \citep{lupu_developing_2016,batalha_color_2018,bailey_nancy_2023}. To image a statistically significant sample of Earth-like planets around Sun-like stars and search for life in their atmospheres, the Astro2020 Decadal survey recommended a UV-optical-IR exoplanet imaging mission, now known as the Habitable Worlds Observatory (HWO) with a $\sim$6.5m coronagraphic space telescope launched in the early 2040s \citep{Decadal2021}. A significant gap in flight-demonstrated starlight suppression still remains between the current state of the art (HST and JWST) and what is needed for Sun-like stars. Lazuli addresses that gap. It uses technologies that are complementary to the ones that Roman is about to fly \citep[e.g.][]{Kasdin2020,cady2025RomanCGIarchitecture} and has similar projected performance with some advantages, due to the 3m telescope, and limitations, due to the limited number of modes; most notably, Lazuli's coronagraph concept omits spectroscopy.

\begin{figure}
\centering
\includegraphics[width=0.99\linewidth]{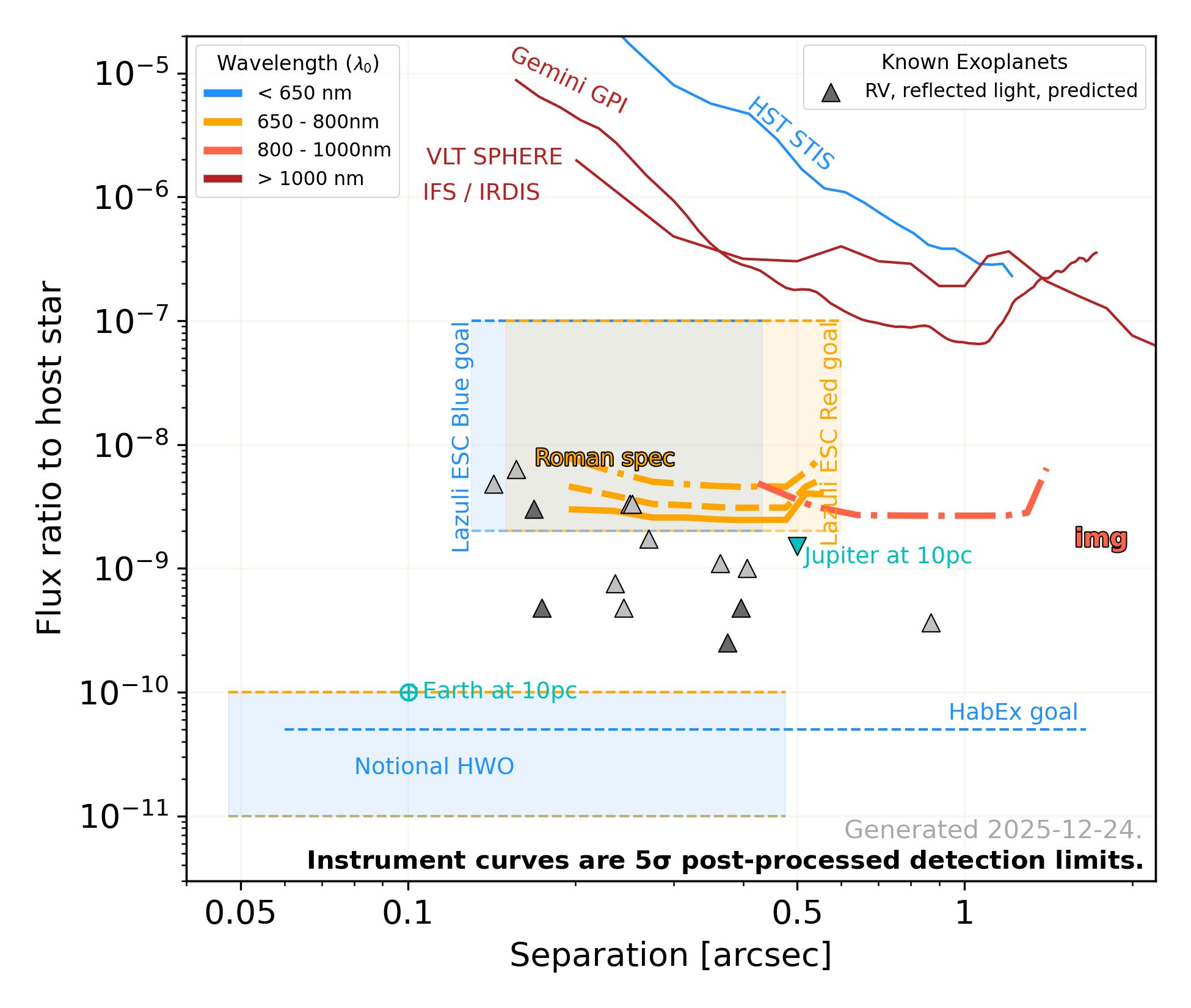}
\caption{Lazuli ESC goal planet-star flux ratio versus distance from the star on the sky compared to Roman and HWO, $5\sigma$ final sensitivity curves adapted from \textit{DI-flux-ratio-plot}\footnote{V. Bailey and S. Hildebrandt, \url{https://github.com/nasavbailey/DI-flux-ratio-plot}}. Compared to Roman, Lazuli's smaller inner working angle goal of 0\farcs12 will enable additional detections and photometry of exoplanets across more of their orbital phase function.  Roman will be able to follow up Lazuli discoveries spectroscopically (dashed yellow lines). HWO's notional sensitivity requirements are shown in the bottom left (blue shaded area, approximately adapted from \cite{stark_paths_2024} and other sources) and reflect two orders of magnitude improvement in post-processed sensitivity. } 
\label{fig:discoveryspace}
\end{figure}

The Lazuli mission's flexibly scheduled, high-throughput $3\;$m-class coronagraphic imaging goal sensitivity of $\leq10^{-8}$ planet-star flux ratios is expected to provide unprecedented detectability of debris disks and giant exoplanets around nearby stars, some of which could be followed up spectroscopically by Roman or HWO (see Figure~\ref{fig:discoveryspace}). These contrast ratios, combined with the expected resolution and throughput of the Lazuli telescope aperture, also enable immediate reconnaissance of the habitable-zones of nearby stars, with a sensitivity commensurate with detections of giant planets and (bright) exozodiacal dust. Detection (or non-detection) of giant planets will provide insights into their occurrence and atmospheric properties, and potentially reveal exomoons \citep{limbach_exomoons_2024,wagner_astrometric_2025}. Further, constraining the orbital locations and architectures of massive planets around the nearest stars will help identify which systems are dynamically compatible with hosting terrestrial planets in the habitable zone \citep[e.g.,][]{kane2025}, which could later be observed with future direct imaging capabilities such as HWO and the Large Interferometer for Exoplanets \citep[LIFE;][]{quanz2022}.

Coronagraph performance depends on stellar magnitude so early searches are expected to be a quick survey of bright stars along with repeated observations of a few cornerstone targets to maximize the multi-visit search completeness around a subset. Figure~\ref{fig:discoveryspace} shows the flux ratio sensitivity expected in context with Roman and HWO. The sensitivity approaches that needed to detect Jupiter analogs around Sun-like stars and systems with known radial velocity planets (upward pointing triangles) are expected to be excellent targets. The Roman Coronagraph's early science is likely to inform Lazuli's target selection strategy. Where Roman and Lazuli overlap, Lazuli will provide shorter wavelength measurements and potentially more complete phase functions closer to host stars. Lazuli's reconnaissance of bright systems will provide new, deep measurements of the scattered light background from exozodiacal dust \citep[e.g.,][]{roberge_exozodiacal_2012,douglas_sensitivity_2022,ertel_review_2025} and presence of giant planets around many potential HWO targets. Future work to maximize the revisit cadence \citep[e.g.,][]{guimond_three_2019,pogorelyuk_deconfusion_2022,bruna_combining_2023}, target selection (e.g. using improved priors or survey optimization with EXOSIMS), filter selection \cite[e.g.,][]{batalha_color_2018}, and develop optimal post-processing and speckle subtraction techniques that leverage onboard telemetry and speckle diversity \citep[e.g.,][]{soummer_detection_2012,amara_pynpoint:_2012,ygouf_data_2015,ren_non-negative_2018,long_more_2024,bonse_use_2025,page_analysis_2025} while following best practices for leveraging artificial intelligence to science data analysis \cite{crilly_ten_2025}.

% ----------------------------------------------
\subsubsection{Exoplanet Transits with the WCC}
\begin{figure}[h!]
\centering
\includegraphics[width=1\columnwidth]{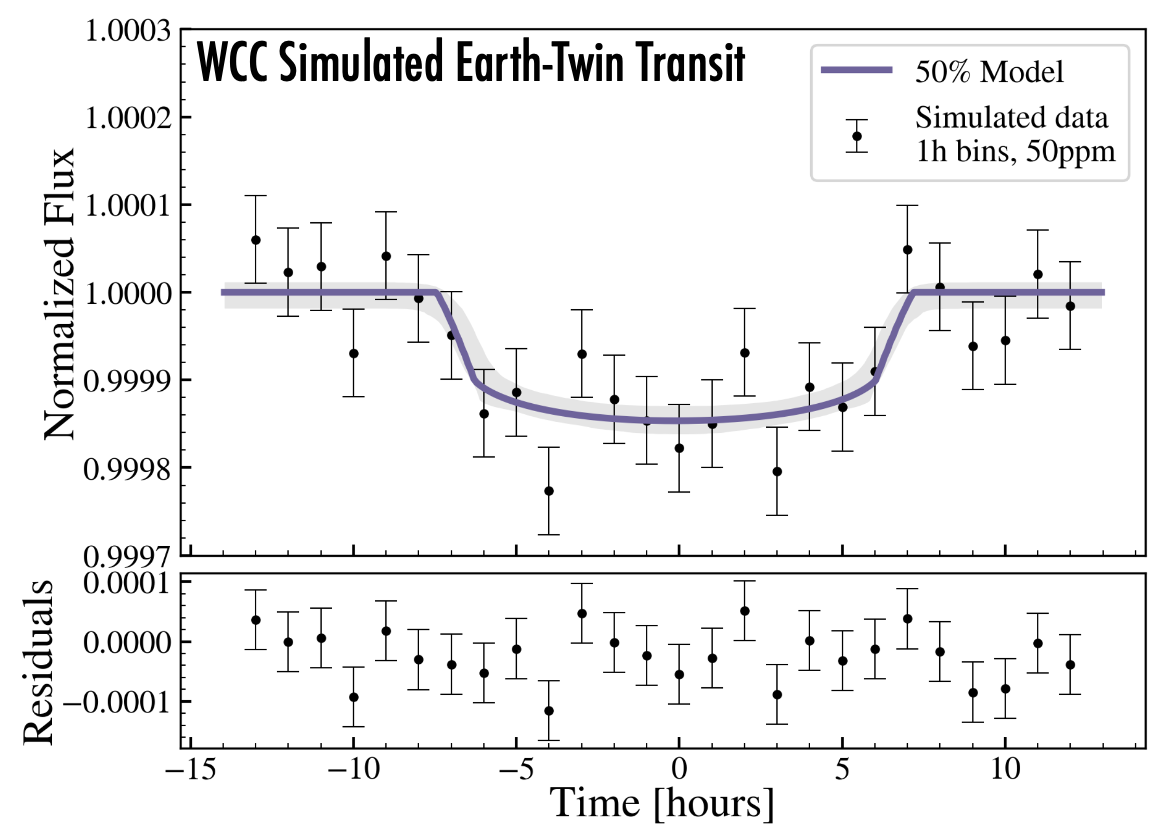}
\caption{Expected transit of an Earth-twin around a sun-like star as observed with the WCC, assuming the WCC achieves its 50ppm precision in 1h effective integration bins observed for two transit durations, or about 26 hours. The median model from a best-fit MCMC simulations (purple line) and corresponding $1\sigma$ credible interval and associated residuals are shown.} 
\label{fig:wcctransit}
\end{figure}

The WCC is being designed to enable the detection and characterization of transiting exoplanets, including Earth analogs—i.e., $\sim$1~$R_\oplus$ planets orbiting within the habitable zones of solar-type stars. An Earth–Sun analog produces a transit depth of $\sim$80~ppm and a transit duration of approximately 13 hours. Achieving a statistically significant detection of such events therefore requires an effective photometric precision of $\sim$50~ppm in one hour of integration (see Figure~\ref{fig:wcctransit}), a performance level that the Lazuli system and the WCC are explicitly aiming to achieve.

To achieve this precision, one WCC sensor will operate in a defocused mode, allowing the stellar point-spread function to be distributed over many pixels. This approach mitigates the impact of inter-pixel sensitivity variations and reduces sensitivity to pointing jitter and guiding errors. As discussed in \S~\ref{sec:wcc}, this sensor will employ a broad, Kepler-like bandpass to maximize photon throughput and thereby minimize photon noise. In addition, the wide field of view of the WCC detectors ensures the presence of multiple nearby reference stars, enabling differential photometry to correct for spacecraft- and detector-related systematics.

This photometric capability enables a broad range of investigations of transiting exoplanets. One application could be a targeted survey of Earth-sized and habitable-zone planets discovered by \textit{Kepler} \citep{borucki2010}, K2 \citep{howell2014}, TESS \citep{ricker2015}, and the upcoming PLATO mission \citep{rauer2014}. Repeated high-precision transit observations can refine orbital ephemerides, improve constraints on planetary radii and densities, and reduce uncertainties in the occurrence rate of terrestrial habitable-zone planets, $\eta_\oplus$, around nearby solar-type stars \citep{fernandez2025challengesInQuantifying,bryson2025etaEarthIsDifficult}. The effectiveness of such a survey in constraining $\eta_\oplus$ will be described in detail in an upcoming publication (Zaman et al. 2026, \textit{in prep.}).

Beyond Earth analogs, the WCC’s photometric precision will enable a wide range of additional exoplanet investigations. These include detailed photometric characterization of high-value transiting systems; the detection of orbital decay through long-baseline, high-precision transit timing measurements \citep[e.g.,][]{patra2017}; searches for transiting exomoons via transit timing variations \citep[e.g.,][]{kipping2009} and/or detection of moon transits \citep[e.g.,][]{teachey2018}; the detection of additional planetary companions through transit timing variations \citep[e.g.,][]{holman2005}; and constraints on planetary obliquities and stellar surface properties through the analysis of starspot-crossing events during transit \citep[e.g.,][]{nutzman2011,sanchisojeda2011}.

\begin{figure*}
\centering
\includegraphics[width=1.0\textwidth]{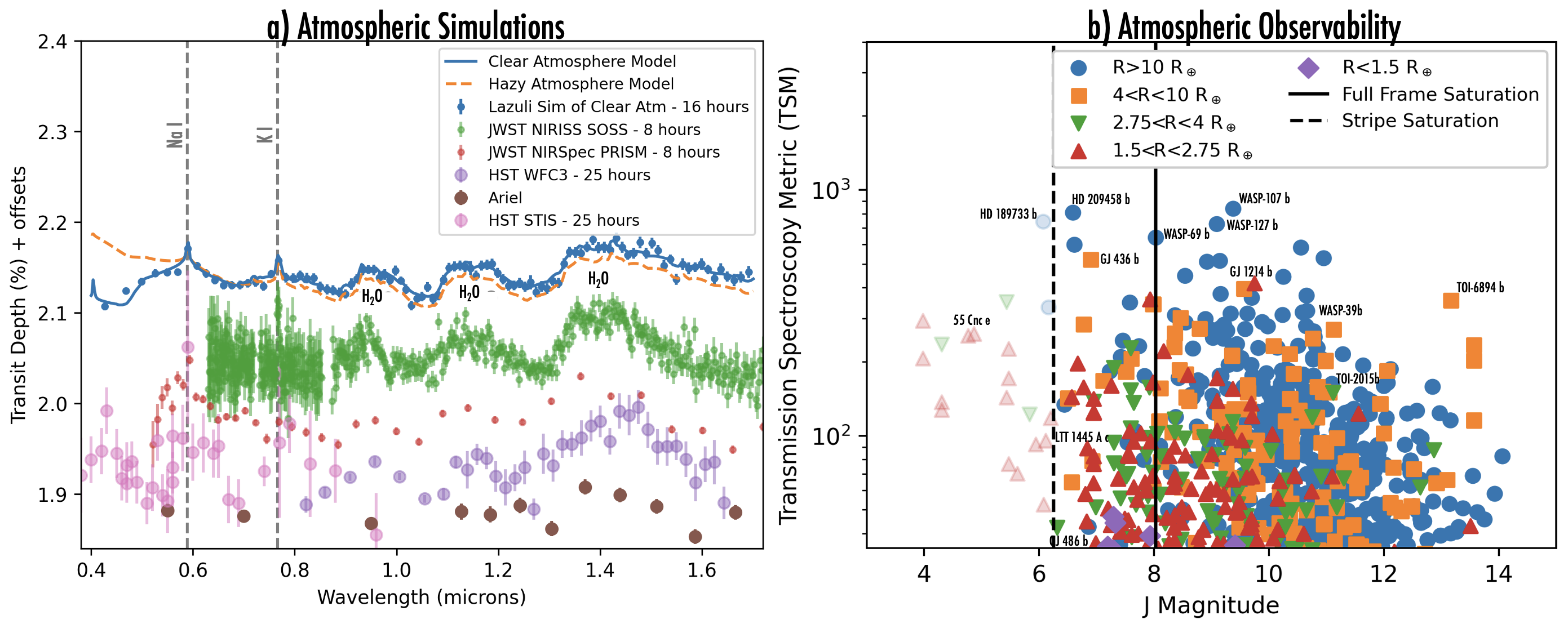}
\caption{a) A simulated transmission spectrum of a WASP-39 b-like exoplanet using \texttt{PICASO} \citep{batalha19refectedLightPICASO} and \texttt{slicersim} \citep{rigault2026slicersim}. The Lazuli IFS spectra (blue data points) span the information-rich optical wavelengths sensitive to hazes, Na and K to the near-infrared water-dominated absorption bands. This complements the infrared capabilities of JWST and Ariel and enables seamless combinations of panchromatic spectra from the visible to near-infrared. b) Exoplanet systems fainter than the subarray saturation limit (dashed vertical line) will be accessible for transmission spectroscopy, including terrestrial, sub-Neptune and giant exoplanets.}
\label{fig:planetatmosphere}
\end{figure*}

% ----------------------------------------------
\subsubsection{Spectroscopy of Transiting Planet Atmospheres}\label{sec:exopAtmospheres}
The Lazuli IFS will have the unique capability to perform space-based high-precision spectroscopy of transiting exoplanets covering the continuous wavelength region from 400--1700~nm. This wavelength region includes key spectroscopic features of Titanium and Vanadium Oxides (TiO, VO), alkali metals (Na, K), water vapor (H$_2$O), methane (CH$_4$) and/or hazes. While many previous ground-based surveys have focused on the alkali elements, and HST and JWST spectroscopic observations have detected alkalis and oxygen-bearing and carbon-bearing molecules in many planets, Lazuli may be the first to simultaneously measure the haze slope, Na, K and water vapor. These combined abundance measurements will constrain the mass-metallicity relation first discovered from the Solar System's carbon abundance \citep[e.g.,][]{atreya2022saturn,kreidberg2014wasp43} to the sodium, potassium, and water only hinted at in previous studies that examine the alkalis and water vapor from different sources \citep[e.g.,][]{welbanks2019massMetalTrends,sun2024massmetallicity}. The Lazuli orbit allows continuous time series spectroscopy without interruptions by Earth eclipses nor day/night temperature swings that can introduce time-dependent systematics and gaps.

Figure~\ref{fig:planetatmosphere} shows a simulation of an exoplanet transmission spectrum of a WASP-39 b-like planet (i.e., same brightness and host star spectrum) using a PICASO version 2.2.1 \citep{batalha19refectedLightPICASO} atmospheric model with no clouds or hazes. We also show a hazy model with a Rayleigh scattering cloud with a reference optical depth of 0.05, wavelength of 250 nm and power law slope of 4.5. We simulate the expected errors for 2 transits of this planet assuming equal in-transit and out-of-transit exposure time. To calculate the expected signal-to-noise, we use the \texttt{slicersim} package version 0.26.0 (Rigault et al. \textit{in prep.}) with default instrument parameters, the narrow field, 12 groups up the ramp with 1 frame per group, a 5400~K, [Fe/H]=0, log(g)=4 host star model \citep{castelli2004models,stsci2013pysynphot} with a J magnitude of 10.67.

Figure~\ref{fig:planetatmosphere} also shows the existing data from JWST NIRISS SOSS \citep{feinstein2023wasp39bNIRISS}, JWST NIRSpec PRISM \citep{rustamkulov2022nirspecPrism}, HST WFC3 \citep{wakeford2018wasp39} and an Ariel simulation from \citet{changeat2025arielSynergies}. While JWST, HST and Ariel all cover the molecular features of hydrogen-bearing and oxygen-bearing molecules, the Lazuli IFS will bridge the visible and near-infrared spectra with a wide simultaneous bandpass. This mitigates against variations from epoch to epoch from stellar activity that can change the transmission spectrum due to the transit light source effect \citep[e.g.,][]{rackham2019lightSourceEffect2}.

Lazuli should have access to a wide variety of planets from small terrestrial planets to giant planets shown in Figure~\ref{fig:planetatmosphere} (panel b). A stripe subarray mode that only reads out a subsection of the detector with all 32 output channels on one row of slicer projections (projections visualized in Figure~\ref{fig:ifs}) will allow observations of targets as bright as J$\approx$6.3 without saturation, depending on the details of the final optical design. This subarray mode will also increase the efficiency of observations near the full frame saturation limit at J$\approx$8.0 from 33\% (2 groups) to 91\% (10 groups). A deeper analysis of Lazuli IFS transmission spectroscopy, including the efficacy of retrievals at various transmission spectroscopy metric thresholds and the resulting accessible exoplanet parameter space, is forthcoming (Pero et al.~2026, \textit{in prep.}).

% ----------------------------------------------
\subsubsection{Characterizing H$\alpha$ Emission from Accreting Protoplanets with the WCC}
The youngest directly imaged exoplanets have been discovered while still embedded within their natal protoplanetary disks. Such ``protoplanets" produce bright H$\alpha$ emission as a consequence of ongoing accretion, making them visible against scattered light from the disks at contrast levels of $\sim 10^{-4}$ (see e.g., \citealt{plunkett2025} and references therein). Comprehensive surveys with ground-based telescopes have hunted for accreting protoplanets in transitional disks \citep{follette2023}, which show dust-depleted gaps and cavities in submillimeter continuum emission \citep{andrews2011} and optical through infrared scattered light \citep{garufi2018}. Since the line-of-sight extinction is lower in the gaps and cavities, H$\alpha$ point sources are expected to be more distinct in these regions \citep{alarcon2024}. However, to date only three sources have been confidently detected, due to challenges with subtracting background disk structure: PDS 70b \citep{keppler2018, wagner2018}, PDS 70c \citep{haffert2019}, and WISPIT 2b \citep{close2025}.

Recent campaigns to observe accreting protoplanets with \emph{HST}/WFC3 have demonstrated the power of using a wide-field camera on a space telescope for high-contrast imaging without a coronagraph \citep{zhou2021, zhou2025}. The improvement in PSF stability from space has also aided in distinguishing between true point source emission and scattered light artifacts from other substructure within the protoplanetary disks \citep{zhou2022, zhou2023}. With this in mind, the Lazuli WCC will carry a narrow-band H$\alpha$ filter, enabling $\sim$0.1-0.3\arcsec~post-processing resolution that matches the radial locations of dust substructures that are resolved in submillimeter emission \citep{andrews_disk_2018, long2018, long2019}. Together with telescope roll angles of $>10^{\circ}$ for space-based angular differential imaging, these capabilities will enable a) characterization of accretion variability from protoplanets detected from the ground (see also \citealt{zhou2025, close2025_variability}) and b) potential surveys to discover new point sources, with target selection guided by ever-increasing theoretical and observational knowledge of disk evolution and radiative transfer \citep{aoyama2018, alarcon2024, cugno2025_extinction}. A detailed exploration of how Lazuli WCC observations can untangle accreting protoplanets from background disk substructure is ongoing (Schneider et al., \textit{in prep.}). 

\subsubsection{Solar System Spectroscopy with the IFS}
Lazuli will also have the capability for non-sidereal tracking, to resolve moving targets within the solar system while minimizing blurring across the detectors. The observatory baselines non-sidereal tracking capabilities of up to 30 mas s$^{-1}$ and a goal of up to 60 mas s$^{-1}$ that are expected to enable observations of the giant planets and their moons, comets, asteroids, Centaurs, and other Kuiper Belt objects \citep{holler2018}. Together with the IFS spectral coverage from 400--1700~nm at $R \sim 100-500$, these capabilities are expected to reveal both water ice and mineral absorption features on targets spanning a wide range of diameters and orbital distances.

% ----------------------------------------------
\subsection{Cosmology}\label{sec:cosmology}
The discovery of the expanding universe initiated the field of observational cosmology, which aims to understand the state, dynamics, and constituents of the universe as traced by astrophysical observables such as luminosity distance and cosmological redshift. The expansion of the universe was discovered using Cepheid variable stars \citep{Leavitt_1912, Hubble_1929}, and the discovery that the expansion is currently accelerating was made using Type Ia supernovae \citep[][SNe~Ia]{perlmutter_measurements_1999, Riess_1998}. Both types of measurements which helped establish the current standard model of cosmology ($\Lambda$CDM), are, with improved statistical uncertainties, now beginning to show surprising evidence for a more complicated model:

The combination of Type~Ia supernovae \citep[\SNeIanospace; ][]{Rubin2025,DESSN5YRCosmology,PantheonPlusCosmology}, baryon acoustic oscillations \citep[BAO][]{DESIY1BAO, DESIDR2BAO} and the cosmic microwave background power spectrum \cite[CMB][]{planck2018} imply an unusual dark energy time variation, while the comparison of the Hubble constant inferred from this CMB measurement with that inferred from local measurements currently disagree \citep{uddin23, riess2024}.

We describe how the capabilities of the Lazuli Space Observatory are well suited to advance our understanding of the universe through observations of Type~Ia supernovae, Cepheid variables and strong gravitational lensing of supernovae.

% ----------------------------------------------
\subsubsection{Type Ia Supernova Cosmology}

Type Ia supernovae have long been prized as cosmological probes due to their high intrinsic luminosity ($M_B\sim-18$~mag) coupled with the ability to standardize their brightnesses either from parameters derived from their lightcurves \citep{Phillips_1993, riess1996, tripp1998} or from their spectra \citep{Fakhouri_2015, Boone_2021b, Stein2022, ganot2025}. Combining these standardized brightnesses with redshifts allows the expansion history of the universe to be measured out to $z\sim2$ \citep[e.g.,][]{DESSN5YRCosmology,Rubin2025}.

The recent surprising indications of time-varying dark energy put a new emphasis on trustworthy supernova distance measurements, since it now becomes particularly important to ensure that both the statistical significance and systematic uncertainties are securely differentiating such time variation from a static dark energy, and then providing reliable indicators of the nature and timing of the variation. For this purpose, it is now possible to employ stronger standardization methods built on spectrophotometry; which while more time-intensive can enhance the major new surveys that discover \SNeIanospace, to accomplish the best currently possible measurements of the expansion history of the universe over the past two thirds of its existence.

The capabilities and timing of the Lazuli IFS will enable it to spectrophotometrically measure \SNeIa discovered by other forefront observatories, including the Roman Space Telescope High Latitude Time Domain Survey (HLTDS) \citep{ROTACReport} and the Vera Rubin Observatory's Deep Drilling Fields (DDF) and Wide Fast Deep (WFD) Survey \citep{LSSTOptimization, LSSTReferenceDesign, RubinCadenceReport3}. Such unified spectrophotometric measurements with broad and uniform wavelength coverage will avoid any discontinuities that might otherwise spring from the cross-calibration of different photometric systems of the discovery surveys, improving the results for the entire community. Figure~\ref{fig:orbit}b shows the visibility of each of the Rubin DDFs and the Roman HLTDS fields throughout one possible year of Lazuli operations. 

These observatories will find transients early enough and with sufficient type-discriminating information that the Lazuli IFS will be able to obtain measurements for a sample having a high purity for \SNeIa near maximum light. This enables the use of spectroscopically ``twin'' \SNeIanospace, a novel technique to standardize \SNeIa using spectroscopy. \cite{Fakhouri_2015}, \cite{Boone_2021b, Boone_2021a}, and \cite{Stein2022} have demonstrated the removal of 3/4ths of the standardized brightness variance using spectrophotometry compared to classical light curve fitting applied to the exact same high-quality data. In addition to the resulting $4\times$ statistical boost for every single supernova, there is a substantial reduction in residuals as a function of host-galaxy environment such as the infamous ``mass step'' \citep{Boone_2021b, ganot2025}. As shown in \cite{Fakhouri_2015}, finding ``twin'' \SNeIa is not hard once the sample size reaches a few hundred, and the method of \cite{Boone_2021b} provides a non-linear 3D latent space that removes the technical need for discrete ``twin'' \SNIanospace.  While the Roman HLTDS will obtain such data in the form of spectral time series using its slitless prism mode \citep[e.g.][]{rubin2025b}, the Lazuli IFS, due to its lower background, larger aperture and focus on \SNeIa at maximum light,  will obtain a substantially larger spectrophotometric sample covering a larger redshift range continuously.

\begin{figure}
\centering
\includegraphics[width=\linewidth]{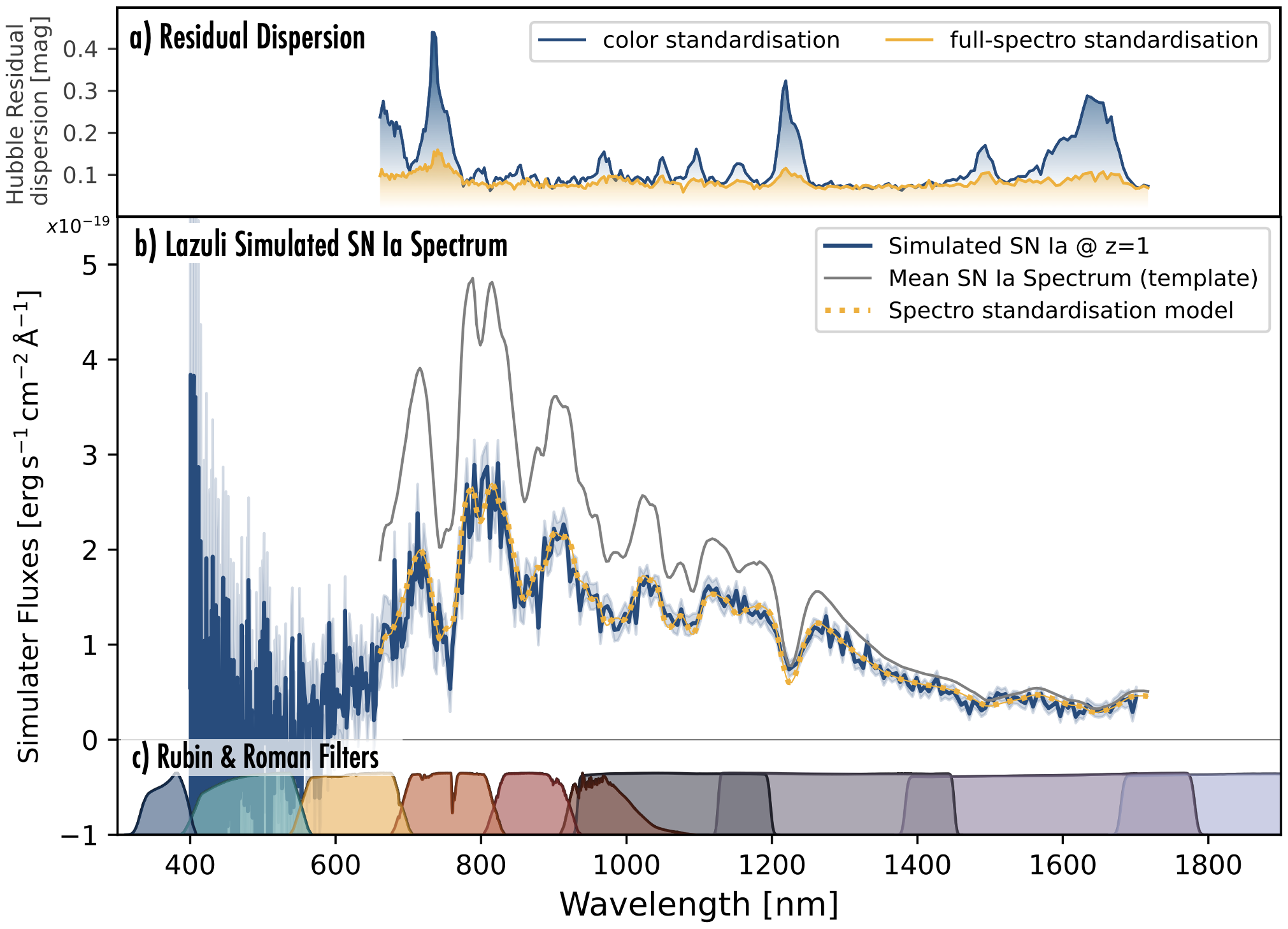}
\caption{Simulated $z=1$ Type Ia supernova spectrum, as observed with the Lazuli IFS in 50~min (panel``b'', in blue) made using \texttt{slicersim} (Rigault et al. 2026). The mean model spectrum is shown in gray, offset above, while the spectroscopic standardization prediction specific to this SN is shown in orange. The amplitude of this model orange line is not a free parameter but derived from the SN spectroscopic behavior \citep{Boone_2021a, Boone_2021b}. Above (panel ``a'') is shown the \SNIa residual brightness scatter for an entire sample after applying the full ``color+3 intrinsic terms'' spectroscopic standardization (orange), or if just using the dust-like color term (blue). This illustrates that the dust-like color term already achieves impressive standardization between \SNIa absorption lines, which strongly vary ($\geq0.3$~mag). Distinguishing these variability origins at every redshift is challenging when employing broad-band filters, as illustrated in the bottom (panel ``c'') for LSST and (SN-related) HLTDS Roman filters.}
\label{fig:snsim}
\end{figure}

Figure~\ref{fig:snsim} illustrates the underpinnings of the spectroscopic standardization approach: given a generic \SNIa spectrum at maximum light, one can reproduce its spectral shape and luminosity given a dust-like color term and the three intrinsic parameters of the \cite{Boone_2021b} non-linear latent space. Figure~\ref{fig:snsim}a illustrates first the spectral variability after removing the dust-like color term (blue line), showing that regions associated with absorption lines have very large residual brightness scatter ($\geq 0.3\,\mathrm{mag}$) while wavelengths in between have little scatter remaining. When next accounting for the terms of the 3D latent space (orange line in Figure~\ref{fig:snsim}a), all wavelengths become well-standardized, leading to a distance modulus scatter of $\sim0.08\,\mathrm{mag}$ \citep{Boone_2021b}; see also discussion in \citep{ganot2025}. Unlike spectroscopy, photometric standardization cannot disentangle these contributions since they cover wavelength ranges narrower than conventional filters. Additionally, stretch and color measured from light curves will be impacted differently as a function of redshift. While the Roman HLTDS will obtain \SNIa spectral time series using its slitless prism mode \citep[e.g.][]{rubin2025b}, the Lazuli IFS, due to its lower background, larger aperture and focus on \SNeIa at maximum light,  will obtain a substantially larger spectrophotometric sample covering a larger redshift range continuously.  

The IFS observer-frame wavelength range from 400--1700~nm allows the observation of the rest-frame wavelength range of 400--680~nm for any target between $z=0$ and $z=1.5$, and the range of 400--850~nm up to $z=1$. This uniquely enables \SNIa spectroscopic standardization based on a common rest-frame window with a single instrument from $z=0$ to $z=1.5$,  as illustrated in Fig.~\ref{fig:snsim}. As already demonstrated by the SEDmachine instrument \citep[SEDm;][]{blagorodnova2018, rigault2019, kulkarni2020}, the Spectrograph for the Rapid Acquisition of Transients \citep[SPRAT;][]{piascik2014}, and the Folded Low Order whYte-pupil Double-dispersed Spectrograph \citep[FLOYDS;][]{brown2013}, the spectral resolution of $100<R<500$ employed by the Lazuli IFS is well suited for rapidly observing transient events like supernovae, which have broad spectroscopic features due to their explosive nature.

Assuming expected performance, the Lazuli IFS will be able to achieve an average S/N of 20 per resolution element for rest-frame wavelengths of 400--680~nm for a typical $z=1$ \SNIa spectrum in a 50~min exposure, as illustrated in Fig.~\ref{fig:snsim}. In comparison, a more nearby target with $z=0.2$ would reach similar S/N levels in a couple of minutes, while a distant $z=1.5$ target would require 4 hours. Increasing the target S/N to 30 per resolution element typically doubles the exposure time. Details concerning the Lazuli IFS exposure time calculator and spectral simulator---called \texttt{slicersim}---will be presented in \citet{rigault2026slicersim}. 

Furthermore, such high-quality spectra will ensure spectroscopic redshifts for all SNe. In cases where there are spectroscopic redshifts for nearby galaxies, this information will aid in selecting the correct host galaxy. For cases with only photometric redshifts, the SN redshifts will reduce the statistical noise and potential systematic biases on the redshift axis of the expansion history measurement \citep[e.g.,][]{rigault2025}.

Because the signal differentiating various cosmology models of interest is small, and since we wish to distinguish them with high confidence, measuring accurate \SNIa fluxes is paramount. Using an IFS rather than a slit spectrograph ensures that all of the SN light is collected, and that observations of the host galaxy after the SN has faded does not rely on expensive pointing accuracy or on a point spread function that is stable over a period of years. Flux calibration of a ground-based IFS to the level afforded by the HST CALSPEC \citep[e.g.,][]{bohlin2020} system has been demonstrated \citep{rubin2022}, as has the accurate subtraction of host galaxy background light \citep{bongard2011}. By comparison, the Lazuli IFS will not need to deal with large variable image quality due to seeing, but diffraction effects, which vary linearly with wavelength, will be very important. \SNeIa at the highest redshifts are faint, requiring large collecting area, high throughput and sufficient spectral and spatial resolution, as well as low detector noise, scattered light and thermal background. The Lazuli IFS is expected to meet these demanding requirements. The parallel fields offer different spatial samplings at comparable spectral resolution, and the fields together cover enough area on the sky for accurate measurement of point spread function wings, sampling of the sky, and host-galaxy subtraction. The need for a linear flux system over a range of $\mathcal{O}(10^4)$ in brightness requires the very precise 2D and 3D calibration systems discussed in \S\ref{sec:ifs}, so that the system can not only provide wavelength and flat-field calibration, but also monitor classical non-linearity, count-rate non-linearity, and problematic pixels in the detector.

Perlmutter et al. \textit{(in prep.)} will provide more detail on a potential design for a powerful \SNIanospace-based study of dark energy behavior over cosmic time that the Lazuli IFS would be capable of conducting with spectrophotometry of $\mathcal{O}(10^4)$ \SNeIanospace. Rather than a two-parameter ($w_0$--$w_a$) fit, the goal would be $\sim$10 redshift bins of luminosity distance measurements out to $z \sim 1.5$. Lazuli spectroscopic follow-up of Roman, Rubin, and other targets from such a survey would also be made available to the community, enabling the (sub)classification of observed transients and determination of redshifts. Separate from the cosmological impact, having a large and uniform supernova sample across a broad range of redshifts can yield key insights into the demographic evolution of \SNeIa and their progenitors or to train photometric classifiers \citep[e.g.,][]{Moller2020,Qu_2021, Burhanudin23,vincenzi2024, deSoto_2024,chen2025}. 

% ----------------------------------------------
\subsubsection{Cepheid Variables and the Hubble Constant}
The Lazuli WCC will have several capabilities uniquely enabling it to contribute to the measurement of extragalactic distances using Cepheid variable stars. Cepheid variables are identified by their characteristic sawtooth-like temporal variations in brightness. In the optical, the amplitudes of these pulsations reach about a factor of two over a $\sim 10$--100d cycle. Twelve epochs of observation in at least one optical bandpass with a power-law sampling is a commonly-used, optimal way to discover Cepheids \citep{Freedman_1994} used since the HST Key Project that resolved the factor of two debate over the value of the Hubble constant \citep{Freedman_2001}. Indeed, the discovery of Cepheids beyond the Local Volume ($d\gtrsim 5$~Mpc) has for 30 years been the exclusive domain of HST, one which Lazuli is poised to join. Note that JWST's large slewing overheads and sharp drop in sensitivity bluewards of 800~nm make the facility far too inefficient for the discovery of Cepheids, though it can still provide high-quality follow-up measurements in the NIR.

The Lazuli telescope will deliver diffraction-limited $r$-band images to the WCC, as well as be capable of fast slews, enabling the efficient discovery of individual Cepheid variables out to at least 40 Mpc. The adoption of SDSS-like bandpasses will enable us to synergize with existing ground-based datasets such as Cepheids in M31 \citep[][PAndromeda]{Kodric_2018} and the upcoming LSST all-sky survey of the southern sky. LSST will provide extremely well sampled observations of Cepheids in all southern targets out to 5~Mpc, providing a definitive calibration of the slope of the Period-luminosity-metallicity (PLZ) relation, which continues to vary well outside of quoted uncertainties, as pointed out by \citet{Majaess_2024, Majaess_2025} and \citet{Hoyt_2025}.

Along with the improved ground-based synergies made possible by our choice of bandpasses, the telescope itself will provide significantly improved optical color measurements of known Cepheid variables in over 30 SN host galaxies, providing more accurate corrections for dust extinction as a result. The existing optical color measurements used by, e.g., \citet{Riess_2022}, for dust corrections are bottlenecked by low S/N, low-cadence observations. Lazuli would also discover new Cepheids in at least 15 more host galaxies that have hosted a SN suitable for cosmology, improving the precision of the calibration of the SN~Ia luminosity. Finally, the flatter QE response of the qCMOS detectors in the WCC focal plane would enable tip of the red giant branch (TRGB) measurements in the $z$-band, which has been demonstrated to be an optimal filter like the $I$-band for accurate TRGB measurements \citep{Bellazini_2024}. An upcoming paper (Hoyt et al. \textit{in prep}) will provide more details on a potential Hubble constant program based on Lazuli and the WCC.

\subsubsection{Strong Gravitational Lensing}

Strong gravitational lensing has long been recognized as a cosmological probe \citep{refsdal1964}, with systematic error sources largely decoupled from either \SNeIa standardized brightnesses or the lower rungs of the distance ladder used to infer the Hubble constant. The most common technique is the measurement of time delays between the different components of a strong lens. Due to their numbers and persistence, such measurements have historically used lensed AGN time delays \citep[e.g., H0LICOW;][]{holicow2020, tdcosmo2025}, which vary over a wide range of timescales in a largely unpredictable manner. Thus, monitoring over the course of years is usually necessary, with specific strong but short-lived variations dominating the time delay signal. More recently monitoring of the transient sky has discovered strongly-lensed supernovae. Though much more rare than AGN, the advantage of supernova lenses is two-fold. First, the variation is comparatively strong and short-lived, offering greater precision for measuring time delays. Second, the SN eventually fades away, allowing the lensing galaxy to be better characterized. Correct measurement of the gravitational potential is the largest source of systematic uncertainty for the time-delay method, so ultimately this advantage is likely to become dominant. One aspect of this issue is the so-called ``mass sheet degeneracy'', which can be broken/reduced when the lensed sources have standardizable luminosities, as with \SNeIanospace. Lazuli spectrophotometry in particular will also offer a spectroscopic means for estimating time delays, and will help account for the effects of microlensing by stars within the lensing galaxy \citep[c.f.,][]{goldstein2018, suyu2024}.

The advent of the Rubin, Roman and LS4 surveys will lead to the discovery of hundreds of gravitationally-lensed supernovae \citep[e.g.,][]{goldstein2019}. The Zwicky Transient Facility has already found several such lensed supernova; one recent example from the literature is the superluminous SN\,2025wny \citep{taubenberger2025, johansson2025} at $z\sim2$ lensed by a pair of galaxies at $z\sim0.4$. Another recent case that is still unfolding is SN\,2025mkn \citep{goobar2025}.

Follow-up of such new gravitationally-lensed SNe will be vigorously pursued by both ground- and space-based facilities. Lazuli's field of regard (cf. Fig.~\ref{fig:orbit}) will allow more temporally-complete monitoring from space of both the key deep fields as well as the wider fields covered by the major imaging surveys. Lazuli's WCC will be able to image these systems in the optical and NIR at spatial scales comparable to JWST\footnote{I.e., Strehl ratios of 0.8 at 633~nm for Lazuli's unobscured 3~m versus 0.8 at 1100~nm for a segmented and obscured 6.5~m (Rigby et al. 2025).}. The Lazuli IFS will be able to classify SN types, provide redshifts, provide better spectroscopic spatial resolution for lens modeling, and deliver spectrophotometric standardization of those that are \SNeIa. {With these types of space-based follow-up, Lazuli will be able to make competitive measurements of the Hubble constant using time delays, as explored recently in, e.g., \cite{suyu2024, hayes2025}.

Lazuli's IFS is also suited to disentangling strong lenses having multiple source planes. The geometry of the system constrains the source distances, while source spectroscopy determines redshifts. This provides a novel way to measure the expansion history of the universe. Since lensed sources can have redshifts of several, this approach can also explore deep into the matter-dominated epoch. An example of an especially beautiful such a system is the ``Carousel Lens,'' with five source planes having been discovered so far \citep{sheu2024}. For this case the MUSE IFS proved specially valuable in identifying the different sources, and the Lazuli IFS can be similarly employed as new such systems are found. Its coverage to bluer wavelengths and with better spatial-sampling than JWST, and higher Strehl ratio than ground-based AO at optical wavelngths, will be especially valuable for identifying Lyman-$\alpha$ emission systems 
% ----------------------------------------------
\section{Mission Operations}
\label{sec:operations}

\subsection{Orbit}
\label{sec:orbit}
Lazuli will operate in a 3:1 lunar-resonant HEO with perigee and apogee altitudes of approximately 70{,}000--285{,}000~km, a 9-day orbital period, and a $29^\circ$ ecliptic inclination. The orbit is selected to maintain a stable resonance with the Moon, in which the spacecraft’s orbital period is a simple integer fraction of the lunar orbital period, resulting in a repeatable long-term geometry. The orbit phasing is chosen such that close lunar perturbations are minimized over the mission lifetime, enabling predictable orbital evolution. This configuration provides a thermally stable environment, a low-radiation regime above Earth’s trapped particle belts, minimal eclipses (approximately 2.4~hours per year), and continuous access to a large fraction of the sky. Near-continuous ground contact enables an average science data downlink of $\sim70$~GB~day$^{-1}$ and rapid response to targets of opportunity within hours of an external trigger.

Several operational orbits were evaluated for Lazuli, including inclined GEO, Sun-Earth L2, Earth-trailing heliocentric, and a range of MEOs. Trade studies examined radiation exposure, eclipse duration and frequency, Earth infrared and albedo effects on instrument thermal stability, downlink data rates versus range, and maneuver complexity for final orbit insertion. These analyses led to the selection of a 3:1 lunar resonant orbit, which provided optimal balance across mission-critical parameters.  

The 3:1 resonance, flown previously by the IBEX mission \citep{IbexMissionPaper}, was selected over the 2:1 resonance flown by TESS because the lower apogee provides approximately 20\% higher downlink capacity while maintaining equivalent sky coverage. Any point on the sky is observable for a minimum of 130 days per year, with continuous viewing zones at ecliptic latitudes $|\beta| \geq 54\degree$ (Figure~\ref{fig:orbit}). Full sky coverage is achieved within 106 days. The orbit is long-term stable, requiring minimal station-keeping maneuvers, maintaining perigee above the geosynchronous belt for at least 100 years, and requiring no end-of-life disposal maneuvers.

Figure~\ref{fig:orbit}a  shows the number of days that Lazuli will be able to view each point on the sky and several fields and targets of interest to potential science cases as described above in \S~\ref{sec:science}. Figure~\ref{fig:orbit}b shows the visibility of each of the fields marked in Figure~\ref{fig:orbit}a for a possible first year of Lazuli operations from June 1, 2028 to June 1, 2029. The Rubin Deep Drilling Fields (DDFs) are visible to Lazuli during their peak period of overhead visibility to Rubin from the ground. The Roman HLTDS fields are continuously visible to both Lazuli and Roman.  Both figures are created using the Ansys / STK (Systems Tool Kit) Access and Coverage modules, integrated with custom python code. As with TESS, there is a continuous viewing zone around the north and south celestial poles. 

\begin{figure*}
\centering
\includegraphics[width=0.99\textwidth]{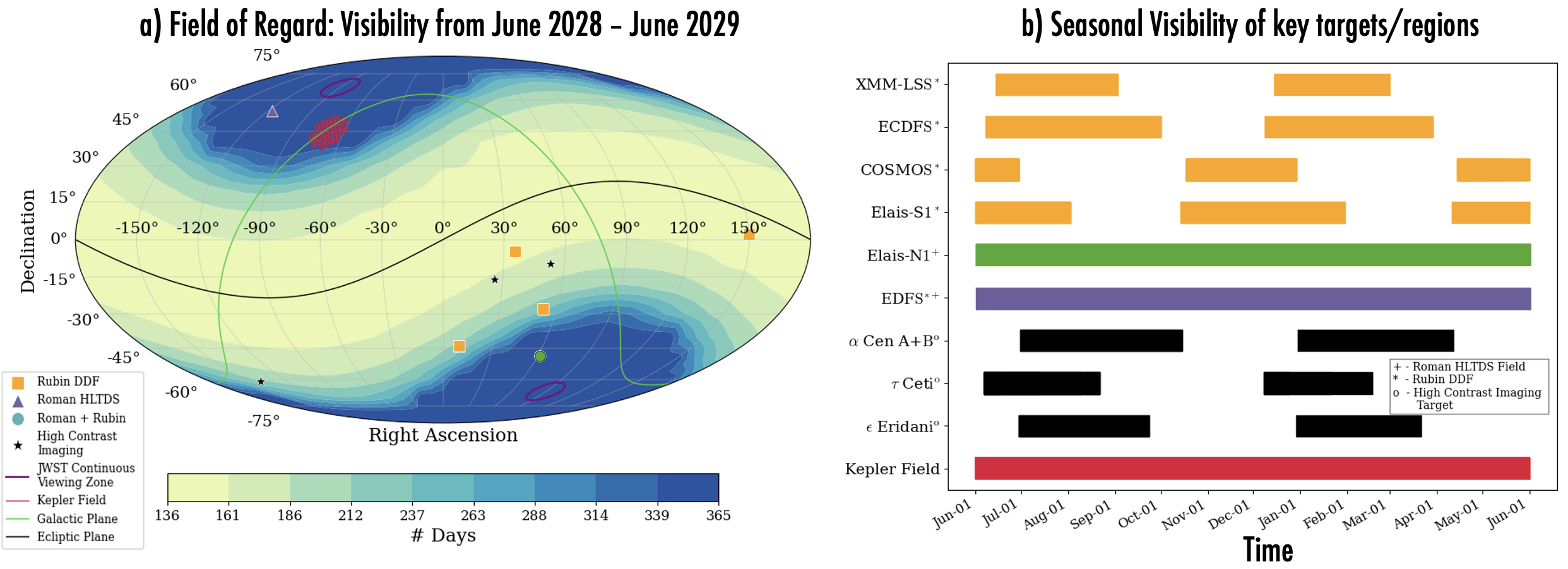}
\caption{a) Field of regard of the Lazuli Space Observatory, showing observability in days (colorbar) across a year as a function of sky position. Key targets and/or fields of interest including the Rubin Deep Drilling Fields (orange squares), Roman HLTDS field (purple triangle) and Roman/Rubin joint fields (green circle), Kepler field (red rectangles), example key high-contrast imaging targets ($\alpha$ Cen, $\tau$ Ceti, $\epsilon$ Eridani; black stars), and the JWST Continuous Viewing Zones (CVZs; purple circles) are highlighted. Additionally, the Galactic plane and Ecliptic plane are included as green and black lines respectively.  b) For each of the key fields plotted in panel a, this panel shows their observability as a function of time throughout one year. Both panels use points sampled on a 10 degree grid over the sky to determine visibility.}
\label{fig:orbit}
\end{figure*}

% ----------------------------------------------
\subsection{Operations Concept}
Lazuli operations are designed around two principles: automation-first execution and rapid response to targets of opportunity. The ground segment will comprise a Science Operations Center (SOC) responsible for science planning, payload commanding, ToO validation, and data processing, and a Mission Operations Center (MOC) responsible for spacecraft bus operations, command uplink, and observatory state-of-health monitoring.

To fulfill a diverse range of science goals—from time-insensitive programs to tightly cadenced monitoring campaigns to disruptive targets of opportunity—all aspects of mission operations are being designed for flexibility and responsiveness. This includes: a dynamic queue scheduling system that can be recomputed on short timescales, balancing ToO interruptions against long-term scheduling efficiency and ongoing program completion; the capability for near-continuous commanding; and a strong emphasis on programmatic decision-making that evaluates both the scientific merit of incoming observations and the cost of disruption to the current schedule. The goal is to begin ToO observations within four hours of trigger receipt (for scientifically justified rapid response ToOs) while maintaining high completion rates for cadenced and baseline programs.

The 4-hour threshold represents a mission requirement rather than a best-case or average figure. The total latency comprises ground-segment processing and schedule regeneration, command generation and uplink, and spacecraft slewing, settling, and target acquisition — the last of which dominates for large slew angles. Under favorable conditions, response times within the 90-minute goal should be achievable. A detailed treatment of the end-to-end ToO workflow and latency budget is presented in \citet{2026arXiv260617136W}.

Central to this approach is the development of an intelligent dynamic queue intended to leverage recent advances in optimization algorithms and machine learning to assess the scientific impact of each observation, balance across multiple programs, and minimize duplication across observatories. The astronomical community has developed a diverse landscape of scheduling approaches---from mixed-integer programming solvers to dynamic figure-of-merit ranking---and Lazuli aims to integrate and build upon these methods. This includes exploring the potential use of large language model agents to augment scheduling decisions, an experimental approach consistent with Lazuli's philosophy of deploying front-line technology with the goal of testing and improving operations for future missions. This lean, automation-driven operations model draws on lessons from large space-based telescopes as well as rapid-response missions such as Swift \citep{Gehrels2004}, and queue-scheduled ground-based facilities including the Hobby-Eberly Telescope \citep{Shetrone2007}, NEID on the WIYN telescope \citep{Schwab2016,schweiker2024}, ESO's Very Large Telescopes (VLT; \citealt{2024SPIE13098E..05A}), and the `Keck Community Cadence' queue for the Keck Planet Finder \citep{Petigura2022}.

\section{COMMUNITY ACCESS \& DATA APPROACH}
\label{sec:community}
The scientific impact of Lazuli will be measured by the excellence of research it enables across the global astronomical community. This section describes the guidelines governing community access to Lazuli observing time, data, and software. The Lazuli Space Observatory is part of the broader Eric and Wendy Schmidt Observatory System, which includes support for science operations, data archiving, and community access as integral parts of the space telescope program. 

\subsection{Community Engagement \& Time Allocation Process}
Engagement with the broader astronomical community is an integral part of the Lazuli mission. Community input is currently incorporated through a set of ‘Science Working Groups' (SWG) aligned with the observatory’s primary capabilities: Time-Domain and Multi-Messenger Astronomy (TDAMM) SWG, Stars and Planets SWG, and Cosmology SWG. Each working group includes external community members and has played a central role in shaping the observatory’s core capabilities, including instrument requirements, observing modes, and performance priorities. As the mission matures and software tools, simulators, and documentation are released, scientists will gain an increasingly concrete understanding of Lazuli’s capabilities and their relevance to specific research areas. In parallel, additional opportunities to contribute are expected to emerge, including engagement through science working groups and contributions to specialized areas such as software, scheduling, instrumentation, or observatory performance management. 

Lazuli is envisioned as a community-access observatory, with observing time expected to be available to the global astronomical community through a merit-based, peer-reviewed Time Allocation Committee (TAC) process. The allocation framework is intended to favor ambitious, collaborative programs that make full use of Lazuli’s unique capabilities, including rapid response, broad wavelength coverage, and stable spectrophotometry, while still accommodating time-critical and disruptive opportunities. Details of the proposal process, allocation cadence, and operational implementation will be finalized as the observatory and its operations concept mature.

\subsection{Data Access \& Release}
\label{sec:DataAccess}
The default posture for Lazuli is open data release without extended proprietary periods. This approach is guided by best practices emerging across the astronomical community and by the scientific case for rapid, multi-facility follow-up—goals that are impeded by extended embargoes, particularly for a mission with a years-long rather than decades-long operational lifetime.

A key consideration in developing data release guidelines is ensuring that open access does not disadvantage proposing teams. To this end, Lazuli is exploring mechanisms to reduce the overhead of proposing---including streamlined submission processes and planning tools---as well as structured support to help awarded investigators move quickly once data are in hand, such as science-ready pipelines, documentation, and analysis tools.

An integrated, cross-observatory Schmidt Observatory System data archive is considered foundational mission infrastructure. Desired features of such an archive include programmatic access via modern APIs, interoperability of data with other astronomical archives, reliable preservation, and---where feasible---co-located compute on a science platform to reduce barriers associated with large data transfers.

\subsection{Software \& Analysis Tools}
Lazuli is conceived as a software-enabled observatory, in which scientific capability is defined not only by hardware performance but by the accessibility, transparency, and extensibility of its data systems. From mission inception, Lazuli’s software ecosystem is being designed to support rapid scientific use, rigorous uncertainty propagation, and community participation.

All mission-developed scientific software—including data reduction pipelines, simulators, exposure time calculators, and archive interfaces—will be released under permissive open-source licenses and maintained in public repositories. This includes instrument-specific pipelines for the WCC, IFS, and ESC, as well as shared infrastructure for calibration handling, metadata validation, and provenance tracking.

The Lazuli pipelines follow a layered data model, progressing from raw, packetized telemetry to calibrated, science-ready products, while preserving intermediate data products and associated metadata to enable independent reprocessing and alternative analysis approaches. Standard community formats are adopted wherever possible (e.g., FITS for images and spectra, Parquet for large catalogs), and pipeline components are designed to be modular rather than monolithic, allowing individual stages to be reused, replaced, or bypassed as scientific needs evolve. 

A defining feature of Lazuli’s software strategy is the tight coupling between simulation, calibration, and analysis. High-fidelity instrument simulators---end-to-end diffraction simulators for the IFS and ESC---along with exposure time calculators such as such as \texttt{slicersim} for the IFS and similar tools for the WCC--are developed alongside the pipelines and share common configuration files and assumptions.

This co-development enables forward-modeling approaches in which detector-level data can be fit directly, preserving photon statistics and correlated noise, while also providing fast “quick-look” reductions for rapid transient classification and follow-up.

Exposure time calculators and performance modeling tools are treated as first-class scientific products rather than ancillary utilities. These tools are version-controlled, scriptable, and designed to interface directly with evolving throughput budgets, calibration knowledge, and mission configuration parameters, enabling reproducible trade studies and transparent assessment of observational feasibility.

Recognizing that software sustainability is essential for scientific impact, the Lazuli project commits to maintaining core analysis tools throughout the mission lifetime, with continuous integration testing, public documentation, and example workflows. Where appropriate, Lazuli will align with and contribute to existing community software ecosystems rather than duplicating effort.

Finally, Lazuli’s software and data systems are explicitly designed to support open science. Data products, pipelines, and simulators are intended to be usable not only by proposing teams but by the broader community immediately upon release, lowering barriers to entry and enabling independent validation, method development, and cross-observatory analyses. In this sense, Lazuli aims not only to deliver data, but to provide a shared computational framework within which new science questions can be posed and answered.

% ----------------------------------------------
% ----------------------------------------------
% ----------------------------------------------
\section{Conclusion}\label{sec:conclusion}
The Lazuli Space Observatory is designed to address a well-defined gap in the current and near-future astrophysical landscape: the absence of a large-aperture, space-based optical–near-infrared facility capable of rapid response, stable spectrophotometry, and broad wavelength coverage. By combining a 3-meter aperture telescope with a focused instrument suite and an operations concept optimized for flexibility, Lazuli enables observations that are difficult or impossible with existing or planned facilities, particularly for fast-evolving and time-critical phenomena.

Lazuli’s capabilities support a broad range of science, spanning time-domain and multi-messenger astronomy, exoplanet characterization, and precision cosmology. Its ability to obtain continuous 400--1700~nm spectrophotometry, multi-band optical imaging, and high-contrast coronagraphic observations from a single platform enables new approaches to transient classification, early-time physical inference, and spectrophotometric standardization. Equally important, the mission is designed to operate in coordination with contemporaneous facilities—including wide-field time-domain surveys, gravitational-wave detectors, and ultraviolet, X-ray, and infrared space observatories—maximizing scientific return through complementary observations.

Beyond its immediate scientific reach, Lazuli serves as a testbed for an alternative model of space observatory development. The mission demonstrates how constrained cost, accelerated schedules, and deliberate risk acceptance can be used to deploy ambitious capabilities while scientific questions remain timely. In this sense, Lazuli functions both as a general-purpose astrophysics facility and as a pathfinder for future missions that prioritize responsiveness, software-enabled operations, and community accessibility.

The capabilities described in this paper represent design requirements and modeled expectations; realizing them fully requires navigating several open challenges. The compressed development timeline, while deliberate, implies a demanding AI\&T schedule and a commissioning period in which all three instruments must be verified to meet their design specifications on orbit. Ground segment development — including data systems, pipeline infrastructure, and community-facing software tools — has necessarily been prioritized behind the space segment and will require sustained effort through launch and into operations. The rapid-response scheduling system, while informed by analysis and heritage from prior missions, has not yet been demonstrated at the system level. These challenges are being actively managed through a program of independent technical reviews and by incorporating best practices from the broader space telescope community, consistent with the mission's philosophy of deliberate risk acceptance in service of accelerated scientific return.

Together, these elements position Lazuli to deliver high-impact science in the late 2020s while informing the design, operation, and scientific use of the next generation of space-based observatories.

\begin{acknowledgments}
This work was supported by Schmidt Sciences. We acknowledge SEE and Quartus for key engineering studies for Lazuli. MR acknowledges, Y. Copin, M. Aubert, and C. Ganot for scientific feedback. Time-domain research by D.J.S. is supported by NSF grants 2108032, 2308181, 2407566, and 2432036, as well as by the Heising-Simons Foundation under grant 20201864. We thank Hiram Olivas and Grant West for their contributions including on the CAD rendering of the ESC in Figure~\ref{fig:escpanel}. 
\end{acknowledgments}

%\software{pysynphot (STScI Development Team 2013)}

\begin{contribution}
AR serves as Mission Scientist and coordinates overall science definition. SF, PK, SP, ESD and SPW contribute to mission leadership. JD, FYY, SRZ, contribute to systems engineering, and observatory implementation. TW, NA, and JL serve as science working group leads. AR, SP, TW, NA, JL, GA, MR, TH, PG, ESD, GF, PI, DK, RA, JH, JE, LL, GS, ES, IJMC, JP, CS, AP, SG, DJS, TM, MAAZ, AS, LP, KK, HC, KD, ME, HGK, DK, KK, JRM, TJM, KM, KLM, PN, JR, SR, SS, IIS, SKS, KVG, AFW, JY, HK and others contribute to science capability definition, instrumentation, optical design, software, data systems, and survey development. All authors contributed to mission development and reviewed the manuscript.
\end{contribution}

\bibliography{bibliography}{}
\bibliographystyle{aasjournal}

\end{document}